\documentclass[12pt]{article}
\usepackage[english]{babel}
\usepackage[latin1]{inputenc}
\usepackage{amsfonts,amssymb,amsmath, epsfig}
\usepackage{color,graphicx,graphics,psfrag}
\usepackage{amsmath,amstext,amssymb,amsfonts, amscd}
\usepackage{hyperref}          

\def\Box{\leavevmode\vbox{\hrule
     \hbox{\vrule\kern4pt\vbox{\kern4pt}%
           \vrule}\hrule}}
\def\blackbox{\leavevmode\vrule height 5pt width 4pt depth 0pt\relax}
\def\endproof{\null\hfill {$\blackbox$}\bigskip}

\newcounter{appendix}
\def\appendix{\advance\c@appendix by 1
   \def\thesection{\Alph{section}}
   \ifnum\c@appendix=1 \setcounter{section}{-1} \fi
   \@startsection {section}{1}{\z@}{-3.5ex plus -1ex minus 
   -.2ex}{2.3ex plus .2ex}{\Large\bf}}

\def\paragraph#1{{\bf #1\ }}

\newtheorem{lemma}{Lemma}[section]  

\newtheorem{theorem}[lemma]{Theorem}

\newtheorem{definition}[lemma]{Definition}

\newtheorem{proposition}[lemma]{Proposition}

\newtheorem{remark}{Remark}[section]

\newtheorem{assumption}{Assumption}[section]

\title{Flow on sweeping networks} 
\author{Pierre Degond $^{(1,2)}$, Michael Herty$^{(3)}$, Jian-Guo Liu$^{(4)}$} 
\date{} 
\begin{document}

\maketitle


\begin{center}
1-Université de Toulouse; UPS, INSA, UT1, UTM ;\\ 
Institut de Mathématiques de Toulouse ; \\
F-31062 Toulouse, France. \\
2-CNRS; Institut de Mathématiques de Toulouse UMR 5219 ;\\ 
F-31062 Toulouse, France.\\
email: pierre.degond@math.univ-toulouse.fr
\end{center}

\begin{center}
3- Department of Mathematics, \\ 
RWTH Aachen University,  D-52062 Aachen,  Germany \\
email: herty@igpm.rwth-aachen.de
\end{center}

\begin{center}
4- Department of Physics and Department of Mathematics\\
Duke University,
Durham, NC 27708, USA\\
email: jliu@phy.duke.edu
\end{center}

\vspace{0.5 cm}
\begin{abstract}
We introduce a cellular automaton model coupled with a transport equation for flows on graphs.  The direction of the flow is described by a switching process where the switching probability dynamically changes according to the value of the transported quantity in the neighboring cells. A motivation is pedestrian dynamics in a small  corridor where the propagation of people in a part of the corridor can be either left or rightgoing. Under the assumptions of propagation of chaos and mean-field limit, we derive a master equation and the corresponding meanfield kinetic and macroscopic models. Steady--states are computed and analyzed analytically and exhibit the possibility of multiple meta-stable states and hysteresis. 
\end{abstract}

\medskip
\noindent
{\bf Acknowledgements:} This work has been supported by KI-Net NSF RNMS grant No. 1107291, grants HE5386/7-1, DAAD 54365630, and the french 'Agence Nationale pour la Recherche (ANR)'  in the frame of the contract 'MOTIMO' (ANR-11-MONU-009-01). MH and JGL are greatful for the opportunity to stay and work at University Paul--Sabatier Toulouse in fall 2012, under sponsorship of Centre National de la Recherche Scientifique and University Paul--Sabatier.


\vskip 0.4cm

\section{Introduction} 
\label{sec:intro}

We are interested in the prediction of qualitative properties and large time behavior of cellular automata (CA) as appearing for example in research on traffic and pedestrian flow \cite{SchadschneiderSeyfried2011aa, Sopasakis2003aa}.  A typical CA is described by a finite set of states lying on a regular lattice and some rules on how to change those within a given time step.  In contrast to existing approaches \cite{Appert-RollandDegondMotsch2011aa, Erban_Haskovec_KRM12, SchadschneiderSeyfried2011aa, Sopasakis2003aa}, we investigate more general graph geometries and we couple this dynamic to a deterministic flow equation for an additional quantity, for example a density. The flow rates in the additional equation depend on the states  of the CA and vice versa. Specifically, we assume that the density sweeps from one cell to one of the neighboring cells according to the state of the CA, hence the terminology of `sweeping network'. On the other hand, the cell-states can switch from one state to another one, according to a probability which depends on an average of the sweeping quantity over the neighboring cells. 

Several examples of applications of such sweeping networks can be envisioned. Our first motivation is the modeling of pedestrian flows in corridors. There, the sweeping quantity is the density of pedestrians in a cell whereas the CA is the ensemble of the cell-states describing in which direction (left or right) pedestrians can move (here in this simple example, we assume that all the subjects in a given cell are forced to move in the direction dictated by the state of the CA but more complex dynamics will be investigated in future work). 

Another example consists of traffic or information networks whose characteristics change with load or occupation. In this case, the sweeping quantity is the load or occupation of the network: it obeys a flow equation whose flow direction is given by the state of the CA at each node. Here, the state of the CA at one node is the index of the neighboring nodes towards which the outgoing flow from the considered node is directed. Hence, the state of the CA does not belong to the set $\{-1,1\}$ but it is still a finite set (which may differ from one node to the next). Therefore, the corresponding dynamical system is not a CA in the restricted sense but shares similar features with CA such as the discreteness of the cell-states. For simplicity, we will still refer to it as a CA.  

One more example is that of supply networks. Such networks are used to describe the flow of parts along a fabrication chain. In some instances, there may exist several suppliers or several clients and the procedure by which the supplier or the client is chosen corresponds to the state of the network. This state may be influenced by the some variables attached to the nodes of the network (such as again, loads, delays, financial reliability, etc.). The choice of the flow direction may in turn determine
the flow of products or money along the network. Further applications include molecular transport in cell biology or bacterial motion \cite{ErbanOthmer2004aa}.

In this paper we present a unified approach to such coupled problems. We consider a CA coupled to a transport equation for a density attached to each cell of the CA where the flux function in the transport is dictated by the CA cell-states locally. Further, the CA cell-states may switch randomly from one value to another one according to a switching rate which depends on an average of the density over the neighboring nodes. We first investigate a simple one-dimensional system, where the nodes are arranged along a line and have a periodic structure. Then, the cell-states are just the variable $z_j \in \{-1,1\}$, where $j$ is the cell index and $z_j = +1$ (resp $z_j = -1$) corresponds to sweeping the density towards the neighboring node to the right (resp. to the left), while the $j$-th cell density is denoted by $\rho_j \in {\mathbb R}_+$.  

From the discrete dynamics, we derive a master equation using a similar presentation as in \cite{CarlenDegondWennberg2011}. The master equation provides the deterministic time evolution of the joint $N$-cell probability distribution function (pdf) ${\mathcal F}(z_1, \ldots z_N, \rho_1, \ldots \rho_N, t)$, where $N$ is the total number of cells of the CA. One distinctive feature of the dynamical systems investigated here lies in the coupling of a stochastic system (the dynamics of the states $z_j$ of the CA) with that of a deterministic system (the transport equation for the cell-densities $\rho_j$). However, the stochasticity of the CA makes the dynamics of the cell-densities random as well. This is why the resulting master equation is posed on the large dimensional space $(z_1, \ldots z_N, \rho_1, \ldots \rho_N) \in \{-1,1\}^N \times {\mathbb R}_+^N$ which encompasses both the cell-state random variables $z_j$ and the cell densities $\rho_j$. This master equation takes the form of a transport equation in the continuous density variables $(\rho_1, \ldots \rho_N)$ and rate equations for the the discrete CA cell-state variables $(z_1, \ldots z_N)$. To our knowledge, this form of a master equation has not been found elsewhere. 

The master equation is posed on a huge dimensional space when $N$ is large and leads to overwhelming numerical complexity for practical use. Additionally, it is difficult to retrieve direct qualitative information, such as analytical solutions, asymptotic behavior of the system, etc., from this complex equation. This is the reason why lower dimensional reductions of this equation are desirable. The goal of this paper is to derive a hierarchy of lower dimensional descriptions of the system. This requires some simplifying assumption, which, in model cases, can be rigorously proven, but which, for the present complex problem, can only be assumed at this stage. 

The first model reduction consists in averaging the $N$-cell pdf over $N-1$ variables, keeping only information on the state of a single cell $j$ by means of the $1$-cell pdf~${\bf f}_j(z_j, \rho_j, t)$. We do not assume cell-indistinguishability so that the $1$-cell pdf of different cells may be different. An equation for ${\bf f}_j$ is easily deduced from the master equation by integrating it over all cell variables $(z_k, \rho_k)$ for $k=1, \ldots, N$ except $k=j$. However, this operation does not lead to a closed equation for ${\bf f}_j$ unless a suitable Ansatz is made for the $N$-cell pdf. This Ansatz is the so-called ``propagation of chaos" which assumes that the cell-states have independent probabilities from each other, i.e. 
$$ {\mathcal F}(z_1, \ldots z_N, \rho_1, \ldots \rho_N, t) \, \,  \approx \, \, \prod_{j=1}^N {\bf f}_j(z_j, \rho_j, t). $$ 
The resulting equation for ${\bf f}_j$ has a similar form as the master equation: it comprises a transport equation in $\rho_j$ and a rate equation for the $z_j$-dependence. But, in contrast to the master equation, it is posed on the low dimensional space $(z_j,\rho_j) \in \{-1,1\} \times {\mathbb R}_+$. Propogation of chaos can be proved in model cases, such as the Boltzmann equation \cite{Lanford_Asterisque76, Mischler_Mouhot_arxiv10, Mischler_etal_arxiv11}, its caricature proposed by Ka\u{c} \cite{Kac_California56} and models of swarming behavior \cite{CarlenDegondWennberg2011, Carlen_etal_PhysD13} (see also \cite{Sznitman_SaintFlour91}). 

The second and last model reduction is to take the limit of an infinite number of cells, i.e. taking the cell-spacing $h$ to zero, while looking at large time-scales, of order $h^{-1}$. This has several consequences. The first one is to legitimate the use of a mean field formula for the switching probabilities for the cell-states. Indeed, as the cell-spacing goes to zero, more and more neighboring cells are included in the computation of the switching probability, leading, through a law of large numbers, to a mean-field evaluation. The second one, related to the change of time-scale is to make the dynamics in $\rho$-space instantaneously convergent to a deterministic dynamics, i.e. the pdf ${\bf f}$ becomes a Dirac delta in $\rho$ at its mean value $\bar \rho(x,t)$ which evolves at the macroscopic time scale according to a classical continuity equation. The flux in this density equation can be expressed in terms of a mean velocity, whose evolution is dictated by an ordinary differential equation derived from the mean-field equation for the switching probabilities. 

The resulting model is a deterministic system of partial differential equation from which all the stochasticity of the original model has disappeared. It bears similarities with the Euler equations of compressible fluid dynamics in that it comprises a continuity equation for the cell density and an evolution equation for the mean velocity. However, there is an important difference in that the velocity equation is a pure ordinary differential equation expressing a relaxation towards a local velocity obtained through some non-local density average. The fact that there is no transport in the velocity equation originates from the fact that the direction of the flux in the sweeping process is purely determined from local quantities at the considered time. Again, we have not found a similar model elsewhere. It is likely though, that adding a time delay in the evaluation of the switching probabilities would restore the spatial transport in the velocity equation. This point will be investigated in future work. 

These general results are then applied to a model of a pedestrian flow. The steady-states of the corresponding fluid model are analyzed. According to the strength of the coupling between the density and the cell-states, we may get multiple steady-states and various kinds of phase transitions (either continuous or discontinuous) between them leading to hysteresis phenomenon. Metastable states and hysteresis are well-documented phenomena in car traffic \cite{Barlovic_etal_EurPhysJB98} and in pedestrian traffic \cite{Helbing_etal_EnvPlanB01}. This allows to establish some qualitative properties analytically. In particular, the occurrence of phase transitions is reminiscent of similar phenomena arising in consensus formation models \cite{Degond_etal_JNonlinearSci13}. The model also bears analogies with the locust model of \cite{Erban_Haskovec_KRM12} but the consideration of cell-states in the present work is original. 

Finally, the presented technique is further refined to treat the case of connected nodes and flows on graphs. Under the propagation of chaos assumption a similar equation for the $1$-cell pdf is obtained. However, the large $N$ limit is not considered because this would necessitate the passage from a discrete network to a continuous space. This limit is outside the scope of the present paper. Still the equation for the discrete $1$-cell pdf is interesting, as it couples the pdf of the neighboring nodes within the flux of the transport term in density space, a feature which we have not observed before.  

CA are widely used models in car traffic \cite{Chertock_etal_M3AS13, Nagel_Schreckenbert_JPhys92, Rickert_etal_PhysA96, SchadschneiderSeyfried2011aa, Sopasakis2003aa, Sopasakis_Katsoulakis_SIAP06} and pedestrian traffic \cite{Blue_Adler_TranspResB01, Burger_etal_KRM11, Burstedde_etal_PhysA01, Nishinari_etal_IEICETransInfSys04}. Standard supply chain models are Discrete Event Simulators \cite{Banks_etal_Prentice99} which bear strong analogies with CA. 

Among Individual-Based models, i.e. models which follow each agent in the course of time, alternatives to CA are particle models such as Follow-the-Leader models in car traffic  \cite{Gazis_etal_OperRes61} and pedestrian traffic \cite{Lemercier_etal_ComputGraphForum12}, or more complex models based on behavioral heuristics \cite{Moussaid_etal_PNAS11}. Kinetic models provide a statistical (and consequently coarser) description of the ensemble of agents. They have been proposed for car traffic in \cite{Prigogine_Herman_1971} and for pedestrian traffic in e.g. \cite{Hoogendoorn_Bovy_TranspResRec07}. Finally, fluid models provide the coarsest - and consequently least computationally intensive - description of traffic systems and has been developed in car traffic in  \cite{Aw_Rascle_SIAP00, Lighthill_Whitham_ProcRoySocA55, Payne_SimulationsCouncilsProcSer71}. They have been recently adapted to pedestrian traffic in  \cite{Appert-RollandDegondMotsch2011aa}. We refer the reader to \cite{Helbing_RevModPhys01} and \cite{Daganzo_LNEconMathSyst03} for reviews on traffic and pedestrian dynamics on the one hand and on supply chain modeling on the other hand. 

The question of proving a rigorous connection between Individual-Based, Kinetic and Fluid models has been treated in e.g. \cite{Aw_etal_SIAP02, Helbing_RevModPhys01} in car traffic, \cite{Degond_etal_arxiv1304.1927, Helbing_ComplexSystems92} in pedestrian traffic and \cite{Armbruster_etal_BullInstMathAcadSinNS07, Degond_Ringhofer_SIAP07} in supply chain modeling. In connection with CA of traffic, it has been investigated in particular in \cite{Chertock_etal_M3AS13, Erban_Haskovec_KRM12}. But, to our knowledge, the present paper provides the first derivation of a fluid model for a CA coupled with the deterministic evolution of a sweeping variable. 

The paper is organized as follows. In section \ref{sec:micro}, we present our sweeping model in one dimension and derive its master equation. In section \ref{sec:kinetic}, we use the propagation of chaos and mean-field assumptions to derive a single-particle closer of the kinetic equation and the hydrodynamic model in the limit of large number of particles and cells. Section \ref{sec:examples} is devoted to an application to pedestrian traffic where meta-stable multiple equilibria and phase transitions are examplified. Section \ref{sec:network} is concerned with the extension of the model to a general graph topology. Finally, section \ref{sec:conclu} provides a conclusion and some perspectives.

\setcounter{equation}{0}
\section{A one-dimensional sweeping model and its master equation} 
\label{sec:micro}

\subsection{The one-dimensional sweeping model}
\label{subsec_oneD}

We are interested in a one-dimensional network consisting of connected cells $j=1,\dots,N$. Each cell contains a certain density $\rho_j \geq 0$ of a given quantity (people, animals, data, goods, particles \ldots) which are able to move or sweep by one cell to the next one. For the simplicity of the presentation, we assume a periodic domain of size equal to $1$, each cell being of size $1/N$. Each cell has a state $z_j \in \{ -1, 1\}$  describing the possible direction of the flow (from left to right  $(z_j=1)$ or from right to left  $(z_j=-1)$). For simplicity we assume that  all particles in cell $j$ move according to the state of the cell $j$ at discrete times $t^n = n \Delta t$, with a time-step $\Delta t$ 	and for $n \in {\mathbb N}$. Hence, the flow of particles $\Psi_{j+\frac{1}{2}}$ across the cell boundary with the $j+1$-th cell is given by 
\begin{equation}
\Psi^n_{j+\frac{1}{2}} = \rho_j^n \max\{ z_j^n, 0 \} +  \rho_{j+1}^n \, \min \{ z_{j+1}^n, 0 \} ,
\label{eq:flux}
\end{equation}	
where the superscript $n$ indicates that the associated quantities are evaluated at time $t^n$. In order to simplify the following discussion, we consider a periodic setting 
$\rho_{j+N}^n = \rho_j^n.$ 

We assume the cell $j$ changing state according to a Poisson process with  rate $\gamma_j^n$ where $\gamma_j^n$ depends on all the cell-states $(z_i^n)_{i=1, \ldots, N}$ and cell-densities $(\rho_i^n)_{i=1, \ldots, N}$.  To be more precise, within a given time interval ${\Delta t}$ the probability to change the state of cell~$j$ is~$1 - \exp \big( -\gamma_j^n \Delta t \big)$, i.e. 
\begin{equation}\label{eq:jumpprocess}
z_j^{n+1 } = z_j^n \zeta_j^n
\end{equation}
where $\zeta_j^n$ is a random variable taking values in $\{-1,1\}$ 
with probability:
\begin{equation}\label{eq:rv}
P(\zeta_j^n = 1) =  e^{-\gamma_j^n \Delta t} \; \mbox{ and } \;  \, P(\zeta_j^n=-1)=1 - e^{-\gamma_j^n \Delta t}. 
\end{equation}
Given some initial data $z_j^0$ and $\rho_j^0$ for $j=1,\dots,N$, the microscopic model for $n\in {\mathbb N}$ is given by 
\begin{equation}\label{eq:micromodel} 
\rho_j^{n+1} = \rho_j^n + N \Delta t( \Psi_{j-\frac{1}{2}}^n - \Psi_{j+\frac{1}{2}}^n ), \quad \rho_{j+N}^n = \rho_j^n.  
\end{equation} 
The factor $N$ in front highlights the fact that the densities change over one time step by an ${\mathcal O}(N \Delta t)$ quantity. This choice is consistant with the choice of the kinetic time scale for the evolution of the cell-states $z^n$ which will be made below. 

We note that the total number of particles is conserved:
$$ \sum_{j=1}^N \rho_j^n = \sum_{j=1}^N \rho_j^0.$$ 
The particle density $\rho_j^n$ is non-negative as soon as the initial density $\rho_j^0$ is so, provided that the time step satisfies the CFL condition $N \Delta t \leq 1/2$. 

\begin{remark}
Many practical networks have finite capacity. This means that the magnitude of the flux is bounded by a maximal value $\Psi^* >0$ and that the expression (\ref{eq:flux}) must be cut-off by this maximal value when it exceeds it. The modifications of the present theory induced by such a cut-off will be discussed in future work. 
\label{rem:capacity}
\end{remark}

We now derive a master equation for this process using the weak formulation as in~\cite{CarlenDegondWennberg2011}. 
Here, the number $N$ of cells will be kept fixed. Later on, we will make $N \to \infty$ in the resulting master equation. 
In a first section, we  derive the master equation for the cell-states, ignoring the dependences of the rates upon the cell densities.

\subsection{A simple cellular automaton for the cell-states and corresponding master equation}
\label{sub:master_cell}

In this section, we first restrict ourselves to the case where the rates $\gamma_j^n$ are independent of the cell-densities $(\rho_i^n)_{i=1, \ldots, N}$. In this case, the dynamics of the cell-states is independent of the cell-densities and the latter can be ignored in the determination of the master equation for the former. Therefore, the random variables are only the states of the cells $z_j^n$ at time $t^n$ and the framework is that of a CA. The discrete state-space at any time for $N$ cells  is therefore $\Sigma^N$ with $\Sigma := \{ \, -1, 1 \}$. We denote by $\vec{z}=(z_i)_{i=1}^{N}$ an element of $\Sigma^N$. A measure $\phi$ on $\Sigma^N$ is defined by the discrete duality with a test function  $\varphi$  as 
\begin{eqnarray}
&&\hspace{-1cm}
\langle \phi, \varphi \rangle_{\Sigma^N} := \sum\limits_{i=1}^N \sum\limits_{ z_i \in \{ -1,1\} }   \phi(\vec{z}) \varphi(\vec{z}).
\label{eq:disc_dual}
\end{eqnarray}
The model is a Markov process. We adapt the classical Markov transition operator formalism to derive the Master equation (see e.g. \cite{CarlenDegondWennberg2011}). The probability distribution function  (pdf) of $\vec{z}$ at time $t^n$ is denoted by  ${\mathcal{F}}^n(\vec{z})$. Let $\varphi$ be any smooth test function on $\Sigma^N$ with values in ${\mathbb R}$ and let ${\mathbb E}$ be the expected value of a random variable. By definition the expectation of the random variable $\varphi (\vec{z}^n)$ for all realizations of $\vec{z}^n$ with distribution ${\mathcal{F}}^n$ is  therefore
\begin{eqnarray}
&&\hspace{-1cm}
{\mathbb E}\left\{ \varphi (\vec{z}^n) \right\} = \langle {\mathcal{F}}^n ,  \varphi \rangle_{\Sigma^N}. 
\label{eq:expect}
\end{eqnarray}
We denote the Markov transition operator from state $\vec{z}^n$ to $\vec{z}^{n+1}$ by $Q$. By definition, it is a bounded non-negativity preserving linear operator on $\ell^2(\Sigma^N)$ with the function $\varphi(\vec{z}) \equiv 1$ as a fixed point. 
The operator $Q$ applied on the test function $\varphi$ from $t^n$ to $t^{n+1}$ and evaluated at state $\vec{z}$ is given by:
\begin{equation}
Q \varphi (\vec{z}) = {\mathbb E} \left\{ \varphi(\vec{z}^{n+1})  \, \big| \,  \vec{z}^n = \vec{z} \right\},
\label{eq:Qphi_def}
\end{equation}
where the expectation is to be taken over all random processes leading from the known state $\vec{z}^n$ to $\vec{z}^{n+1}$.  
Hence, $Q\varphi(\vec{z}^n)$ is a random variable for all realizations of $\vec{z}^n$ with distribution ${\mathcal{F}}^n$. Therefore, its expectation is  
\begin{align*} \, 
{\mathbb E} \left\{ Q\varphi(\vec{z}^n) \right\} &= \langle {\mathcal{F}}^n,  Q \varphi \rangle_{\Sigma^N} =  \langle Q^* {\mathcal{F}}^n, \varphi  \rangle_{\Sigma^N} ,
\end{align*}
where $Q^*$ is the $\ell^2$ adjoint operator to $Q$. Due to the property of the conditional expectation, we also have
\begin{align*} \, 
{\mathbb E} \left\{ Q\varphi(\vec{z}^n) \right\} &=
{\mathbb E} \left\{ {\mathbb E} \left\{ \varphi(\vec{z}^{n+1})  \, \big| \, \vec{z}^n \right\}   \right\} = 
{\mathbb E} \left\{ \varphi( \vec{z}^{n+1}) \right\} 
=  \langle {\mathcal{F}}^{n+1} , \varphi \rangle_{\Sigma^N} .
\end{align*}
Combining and noting that the previous equations hold for all functions $\varphi$, we have:
\begin{eqnarray}
{\mathcal F}^{n+1}( \vec{z} ) \, = Q^* {\mathcal F}^n(\vec{z}).
\label{eq:meanfieldderv_1}
\end{eqnarray}
We will show that $(Q^* - \mbox{Id}) {\mathcal F} = O(N \Delta t)$. Therefore, the rate of change of the pdf over one time-step is 
\begin{eqnarray*}
\frac{{\mathcal F}^{n+1} - {\mathcal F}^{n}}{N \Delta t} ( \vec{z} ) \, = \frac{1}{N \Delta t} (Q^* - \mbox{Id}) {\mathcal F}^n(\vec{z}) = O(1).
\end{eqnarray*}
In the limit $N \Delta t \to 0$, with $n N \Delta t \to t$, we have ${\mathcal F}^{n} (\vec z) \to {\mathcal F} (\vec z,t)$ with
\begin{eqnarray}
\label{eq:meanfieldderv_2}
\frac{\partial {\mathcal F}}{\partial t} ( \vec{z},t ) \, = \lim_{N \Delta t \to 0} \frac{1}{N \Delta t}  (Q^* - \mbox{Id}) {\mathcal F}( \vec{z},t ) = L^* {\mathcal F}( \vec{z},t ).
\end{eqnarray}
This is the so-called time-continuous master equation of the process and the operator $L$ (the adjoint to $L^*$) is called the Markov generator. This choice of time scale is called the kinetic time scale. It corresponds to each particle colliding in the average once during one time step $\Delta t$.

\begin{proposition}
The master equation for the time-continuous version of the CA described in section \ref{subsec_oneD} when the rates $\gamma_j$ are independent of the cell-densities $(\rho_i)_{i=1, \ldots, N}$ is given, at the kinetic time-scale, by 
\begin{equation}
\frac{\partial}{\partial t}  {\mathcal{F}} (\vec{z}, t)  = \frac{1}{N} \,  \sum_{j=1}^N  ( \gamma_j(-z_j, \hat z_j) \mathcal{F}(-z_j, \hat z_j, t)  - \gamma_j(z_j, \hat z_j) \mathcal{F}(z_j, \hat z_j, t) ).    
\label{eq:master_CA}
\end{equation}
where ${\mathcal{F}} (\vec{z}, t)$ is the time-continuous joint pdf of the cell-states and where we denote by $\hat z_j = (z_1, \ldots , z_{j-1} , z_{j+1}, \ldots , z_N)$ and for a function $\phi( \vec z)$, $\phi(z_j, \hat z_j) = \phi (\vec z)$ and $\phi(-z_j, \hat z_j) = \phi (z_1, \ldots , z_{j-1} , - z_j ,  z_{j+1}, \ldots , z_N)$. 
\label{prop:master_CA}
\end{proposition}

The operator at the right-hand side of (\ref{eq:master_CA}) contains two terms. The first term is positive and describes the increase of the pdf due to cells which reach the state $\vec z$ after switching from a different state (namely from the $j$-th cell state $-z_j$). The increase occurs at rate $\gamma_j(-z_j, \hat z_j)$. The second term is negative and describes the decrease of the pdf due to cells which leave the state $\vec z$ for a different one (namely the $j$-th cell state $z_j$). The decrease occurs with rate $\gamma_j(z_j, \hat z_j)$. The resulting expression has to be summed up over all possible cells $j \in [1,N]$. The weighting factor $\frac{1}{N}$ is there to ensure that the proper time scale has been chosen to ensure the finiteness of the right-hand side in the limit $N \to \infty$. This is the so-called kinetic time scale, where, on average, a given cell changes state only a finite number of times over a finite time interval. 

\medskip
{\bf Proof of Proposition \ref{prop:master_CA}.}
Let $\varphi$ be a smooth test function. We have: 
\begin{eqnarray}
&&\hspace{-1cm} 
 \langle (Q^* - \mbox{Id}) {\mathcal F}^{n}, \varphi \rangle_{\Sigma^N}
= {\mathbb E} \left\{ {\mathbb E} \left\{ \varphi(\vec{z}^{n+1})- \varphi(\vec{z}^{n}) \, \big| \,  \vec{z}^n \right\}  \vec{z}^n \right\} \nonumber\\
&&\hspace{-1cm} 
=  \big\langle \, \mathcal{F}^{n}(\vec{z}) ,  \sum_{j=1}^N (\varphi(-z_j, \hat z_j) - \varphi(z_j, \hat z_j) ) (1 - e^{- \gamma_j(\vec{z})\Delta t}) \prod_{i \not = j} e^{- \gamma_i(\vec{z}) \Delta t} \, \,  \big\rangle_{\Sigma^N} \nonumber \\ 
&&\hspace{10cm} 
+ O((N\Delta t)^2)
\label{eq:master1}
\\
&&\hspace{-1cm} 
= \Delta t \,   \sum_{j=1}^N  \langle \,\mathcal{F}^{n}(z_j, \hat z_j) , (\varphi(-z_j, \hat z_j) - \varphi(z_j, \hat z_j) ) \, \gamma_j(z_j, \hat z_j) \,  \rangle_{\Sigma^N} + O((N\Delta t)^2) \nonumber\\
&&\hspace{-1cm} 
=  \Delta t \,  \langle \, \sum_{j=1}^N \{ \gamma_j(-z_j, \hat z_j) \mathcal{F}^{n}(-z_j, \hat z_j) - \gamma_j(z_j, \hat z_j)  \mathcal{F}^{n}(z_j, \hat z_j) \} , \varphi(\vec{z}) \, \rangle_{\Sigma^N}  + O((N\Delta t)^2).\nonumber
\end{eqnarray}
To derive (\ref{eq:master1}), we note that the probability that a given $k$-tuple of cells switch states is $O\big(\Delta t^k\big)$ but there are $O(N^k)$ possible $k$-tuple of cells. Hence, the total probability that $k$ cells change is $O((N \Delta t)^k)$. Therefore, the probability that there are strictly more than one change is $O((N \Delta t)^2)$ while that of only one change is $O( N \Delta t)$. We note that the probability of no change is dropped out by the subtraction. Then, we have: 
$$
  \frac{ {\mathcal{F}}^{n+1} (\vec{z})  -\mathcal{F}^{n} (\vec{z}) }  { N \Delta t}
   =  \frac{1}{N} \, \sum_{j=1}^N  ( \gamma_j(-z_j, \hat z_j) \mathcal{F}^{n}(-z_j, \hat z_j) 
- \gamma_j(z_j, \hat z_j) \mathcal{F}^{n}(z_j, \hat z_j) )   + O(N \Delta t), 
$$
and, in the limit $N \Delta t \to 0$, we get (\ref{eq:master_CA}). \endproof

In the next section, we consider the full process where the rates $\gamma_j^n$ depend on the cell-densities $(\rho_i^n)_{i=1, \ldots, N}$.

\subsection{The master equation for the sweeping process}
\label{sub:master_dens}

We now consider the full sweeping process as described in section \ref{sec:micro}. 
The random variables are now the states of the cells $z_i \in \{ \, -1, 1 \} $ and the number of particles within each cell $\rho_i \in {\mathbb R}_+$. The discrete state space for  $N$ cells  is therefore ${\mathbb A}^N$ with ${\mathbb A} := \{ \, -1, 1 \}  \times {\mathbb R}_+$. We still denote by $\vec{z}=(z_i)_{i=1}^{N}$ and similarly for $\vec{\rho}=(\rho_i)_{i=1}^{N}$. A measure $\phi$ on ${\mathbb A}^N$ is defined by its action on a continuous function $\varphi$ on ${\mathbb A}^N$ by: 
\begin{eqnarray*}
\langle \phi , \varphi \rangle_{{\mathbb A}^N} =  \sum_{ \vec{z} \in \{ -1,1\}^N } \int_{\vec{\rho} \in {\mathbb R}_+^N}  \phi(\vec{z},\vec{\rho}) \, \varphi(\vec{z},\vec{\rho}) \, d \rho_1 \ldots d \rho_N.
\end{eqnarray*}
We also denote $\gamma_i=\gamma_i(z_i,{\hat{z}}_i,\vec{\rho})$, $\vec{\Psi}_+ = (\Psi_{1+\frac{1}{2}}, \ldots , \Psi_{i+\frac{1}{2}}, \ldots , \Psi_{N+\frac{1}{2}})$,  $\vec{\Psi}_- = (\Psi_{1-\frac{1}{2}}, \ldots , $ $\Psi_{i-\frac{1}{2}},\ldots , \Psi_{N-\frac{1}{2}})$. Then, the vector version of the density update is 
\begin{equation} 
\vec{\rho}^{n+1} - \vec{\rho}^n + N \Delta t \, ( \vec{\Psi}_+^n - \vec{\Psi}_-^n ) =0. 
\label{eq:rho_update}
\end{equation}

\begin{proposition}
The master equation for the time-continuous version of the sweeping process described in section \ref{subsec_oneD} when the rates $\gamma_j$ depend on both the cell-states $(z_i)_{i=1, \ldots, N}$ and  the cell-densities $(\rho_i)_{i=1, \ldots, N}$ is given, at the kinetic time-scale, by 
\begin{eqnarray}
&&\hspace{-1cm}
\Big( \frac{\partial}{\partial t}  {\mathcal{F}}  
-  \nabla_{\vec \rho} \cdot \big( \big( \vec{\Psi}_+ - \vec{\Psi}_-\big) \, {\mathcal F} \big) \Big) (\vec{z}, {\vec \rho}, t)   \nonumber \\
&&\hspace{0cm}
=   \frac{1}{N} \, \sum_{j=1}^N  ( \gamma_j(-z_j, \hat z_j, {\vec \rho},t) \mathcal{F}(-z_j, \hat z_j, {\vec \rho}, t) 
- \gamma_j(z_j, \hat z_j, {\vec \rho},t) \mathcal{F}(z_j, \hat z_j, {\vec \rho}, t) )   ,
\label{eq:master8}
\end{eqnarray}
in strong form or 
\begin{eqnarray}
&&\hspace{-1cm}
\langle \frac{\partial {\mathcal F}}{\partial t}, \varphi \rangle_{{\mathbb A}^N} 
=  - \langle {\mathcal F} , \nabla_{\vec \rho} \varphi \cdot \big( \vec{\Psi}_+ - \vec{\Psi}_-\big) \, \rangle_{{\mathbb A}^N} \nonumber \\
&&\hspace{1cm}
+  \frac{1}{N} \,  \sum_{j=1}^N \big \langle \,  \mathcal{F}(z_j, \hat z_j, {\vec \rho}) , \gamma_j(z_j, \hat z_j, {\vec \rho}) \,  \{ \varphi(-z_j, \hat z_j, {\vec \rho}) - \varphi(\vec{z}, {\vec \rho}) \} \, \big\rangle_{{\mathbb A}^N} , 
\label{eq:master9}
\end{eqnarray}
for any smooth test function $\varphi$ on ${\mathbb A}^N$ with values in ${\mathbb R}$, in weak form. We have noted $ \nabla_{\vec \rho} \varphi \cdot \vec g = \sum_{j=1}^N g_j \partial_{\rho_j} \varphi $ and $\nabla_{\vec \rho} \cdot \vec g \, \varphi =  \varphi \sum_{j=1}^N \partial_{\rho_j} g_j$ for any functions $\varphi(\vec\rho)$ and $\vec g (\vec \rho) = (g_j(\vec \rho))_{j=1, \ldots , N}$. 
\label{prop:master_full}
\end{proposition}

The right-hand side of (\ref{eq:master8}) has the same structure as that of (\ref{eq:master_CA}). We refer the reader to the paragraph following Prop. \ref{prop:master_CA} for its interpretation.  The time-derivative at the left-hand side is now supplemented with a first order differential term in $\vec{\rho}$ space (the second term). Due to (\ref{eq:flux}), the coefficient $\big( \vec{\Psi}_+ - \vec{\Psi}_-\big)$ inside this derivative couples the neighboring nodes of each cell $j$. It expresses how the density evolves as a consequence of the density in cell $j$ sweeping to one of its neighboring cells, and the density in the neighboring cells sweeping into the $j$-th cell. Because the stochasticity of the dynamics of the cell-states $z_j$ is propagated to the densities $\rho_j$, the description of the densities is through the pdf ${\mathcal F}$. Therefore, the density evolution translates into a transport equation in density space for the pdf.

\medskip
{\bf Proof:} Let $\varphi$ be any smooth test function on ${\mathbb A}^N$ with values in ${\mathbb R}$. We write
\begin{eqnarray}
&&\hspace{-1cm}
\langle {\mathcal{F}}^{n+1} -\mathcal{F}^{n}, \varphi \rangle_{{\mathbb A}^N}
= {\mathbb E} \left\{ {\mathbb E} \left\{ \varphi(\vec{z}^{n+1}, {\vec \rho}^{n+1})- \varphi(\vec{z}^{n}, {\vec \rho}^{n}) \, \big| \,  (\vec{z}^n,{\vec \rho}^n) \right\}  (\vec{z}^n,{\vec \rho}^n) \right\} \nonumber \\
&&\hspace{0cm}
= {\mathbb E} \left\{ {\mathbb E} \left\{ \varphi(\vec{z}^{n+1}, {\vec \rho}^{n+1})
-\varphi(\vec{z}^{n+1}, {\vec \rho}^{n})  \, \big| \,  (\vec{z}^n,{\vec \rho}^n)  \right\} (\vec{z}^n,{\vec \rho}^n)  \right\} \nonumber  \\
&&\hspace{3cm}
+ {\mathbb E} \left\{ {\mathbb E} \left\{  \varphi(\vec{z}^{n+1}, {\vec \rho}^{n})
- \varphi(\vec{z}^{n}, {\vec \rho}^{n}) \, \big| \,  (\vec{z}^n,{\vec \rho}^n)  \right\}  (\vec{z}^n,{\vec \rho}^n)  \right\} \nonumber \\
&&\hspace{0cm}
=  I + II, \label{eq:master0}
\end{eqnarray}
together with (\ref{eq:rho_update}). Then, we have:
\begin{eqnarray}
I
&=& 
{\mathbb E} \Big\{ \, {\mathbb E} \big\{  \nabla_{\vec \rho}\varphi(\vec{z}^{n+1}, {\vec \rho}^{n})({\vec \rho}^{n+1}-{\vec \rho}^{n}) \, \big| \,  (\vec{z}^n,{\vec \rho}^n)  \big\} \, \,  (\vec{z}^n,{\vec \rho}^n)  \Big\} + {\mathcal O}((N \Delta t)^2) \nonumber \\
&=& 
- N \Delta t \, {\mathbb E} \Big\{ \, {\mathbb E} \big\{ \nabla_{\vec \rho} \varphi(\vec{z}^{n+1}, {\vec \rho}^{n}) ( \vec{\Psi}_+^n - \vec{\Psi}_-^n)  \, \big| \,  (\vec{z}^n,{\vec \rho}^n)  \big\} \, \,  (\vec{z}^n,{\vec \rho}^n)  \Big\} + {\mathcal O}((N \Delta t)^2) \nonumber \\
&=& 
- N \Delta t \, {\mathbb E} \Big\{ \, {\mathbb E} \big\{ \nabla_{\vec \rho} \varphi(\vec{z}^{n}, {\vec \rho}^{n}) ( \vec{\Psi}_+^n - \vec{\Psi}_-^n)  \, \big| \,  (\vec{z}^n,{\vec \rho}^n)  \big\} \, \,  (\vec{z}^n,{\vec \rho}^n)  \Big\} + {\mathcal O}((N \Delta t)^2) ,
\label{eq:master2}
\end{eqnarray}
where for the second equality, we have used (\ref{eq:rho_update}). For the third one, we note that the probability for no state change is given by $\prod_{j=1}^N \exp (-\gamma_j^n \Delta t) = 1 - {\mathcal O}(N \Delta t)$ and therefore, the probability for at least one change is ${\mathcal O}(N \Delta t)$. Then, we can remove the inner expectation in (\ref{eq:master2}) because there is no change involved. Using the definition of the outer expectation, we can recast (\ref{eq:master2}) as follows: 
\begin{eqnarray}
I &=& - N \Delta t \, \langle {\mathcal F}^n , \nabla_{\vec \rho} \varphi \cdot ( \vec{\Psi}_+ - \vec{\Psi}_-)  \, \rangle_{{\mathbb A}^N} + {\mathcal O}((N \Delta t)^2) \label{eq:master3}
\\
&=& N \Delta t \, \langle \nabla_{\vec \rho} \cdot \big( ( \vec{\Psi}_+ - \vec{\Psi}_-) \, {\mathcal F}^n \big) ,  \varphi \rangle_{{\mathbb A}^N}  + {\mathcal O}((N \Delta t)^2) . \label{eq:master3.1} 
\end{eqnarray}

For the second term, the algebra proceeds exactly like in section \ref{sub:master_cell}. Details are omitted. As an outcome we get: 
\begin{eqnarray}
&& \hspace{-1.5cm}
II = \Delta t \,  \sum_{j=1}^N \big \langle \,  \mathcal{F}^{n}(z_j, \hat z_j, {\vec \rho}) , \gamma_j(z_j, \hat z_j, {\vec \rho}) \,  \{ \varphi(-z_j, \hat z_j, {\vec \rho}) - \varphi(\vec{z}, {\vec \rho}) \} \, \big\rangle_{{\mathbb A}^N}  + {\mathcal O}((N \Delta t)^2) \label{eq:master4}\\
&& \hspace{-1cm}
=  \Delta t \, \big \langle \,  \sum_{j=1}^N  \{ \gamma_j(-z_j, \hat z_j, {\vec \rho}) \mathcal{F}^{n}(-z_j, \hat z_j, {\vec \rho}) 
- \gamma_j(z_j, \hat z_j, {\vec \rho}) \mathcal{F}^{n}(z_j, \hat z_j, {\vec \rho}) \} , \varphi(\vec{z}, {\vec \rho}) \, \big\rangle_{{\mathbb A}^N} \nonumber \\
&& \hspace{10cm}
+ {\mathcal O}((N \Delta t)^2) . 
\label{eq:master4.1}
\end{eqnarray}
Inserting (\ref{eq:master3.1}) and (\ref{eq:master4.1}) into (\ref{eq:master0}) leads to: 
\begin{eqnarray}
&&\hspace{-1cm}
\langle \frac{{\mathcal F}^{n+1} -{\mathcal F}^{n}}{N \Delta t}, \varphi \rangle_{{\mathbb A}^N} 
=  - \langle {\mathcal F}^n , \nabla_{\vec \rho} \varphi \cdot ( \vec{\Psi}_+ - \vec{\Psi}_-)  \, \rangle_{{\mathbb A}^N} \nonumber \\
&&\hspace{-1cm}
+  \frac{1}{N} \sum_{j=1}^N  \big \langle \,  \mathcal{F}^{n}(z_j, \hat z_j, {\vec \rho}) , \gamma_j(z_j, \hat z_j, {\vec \rho}) \,   \{ \varphi(-z_j, \hat z_j, {\vec \rho}) - \varphi(\vec{z}, {\vec \rho}) \} \, \big\rangle_{{\mathbb A}^N} 
+ {\mathcal O}(N \Delta t) . 
\label{eq:master5}
\end{eqnarray}
Now, letting $N \Delta t \to 0$ in (\ref{eq:master5}) and $n N \Delta t \to t$, we find the weak form (\ref{eq:master9}) of the master equation. Then, since the test function $\varphi$ is arbitrary, using (\ref{eq:master3}), (\ref{eq:master4}) and the same $\Delta t \to 0$ limit and passage to the kinetic time-scale as for the weak form, we get the strong form (\ref{eq:master8}) of the master equation. \endproof

\setcounter{equation}{0}
\section{Single-particle closure and macroscopic model}
\label{sec:kinetic}

\subsection{Goal}
\label{subsec:goal}

The description of the system by means of the $N$-cell pdf is too complicated and cannot be practically used, neither numerically nor analytically. The goal of this section is to propose a reduction of the system to a $1$-cell pdf (i.e. the $1$-cell marginal of the pdf ${\mathcal{F}}(\vec{z},{\vec \rho},t)$), and to compute its time evolution. A straightforward integration of the master equation does not lead to a closed equation for the $1$-cell pdf. The goal of this section is to propose a closure of this equations by assuming that propagation of chaos holds. Then, we investigate the limit of $N \to \infty$ and postulate that the rates can be approximated by mean-field approximation. In this limit, we find a system of hydrodynamic equations. 

We first define the marginals of the pdf as follows: 

\begin{definition}
For any $j \in \{1,2\dots, N \}$, we define the 
marginal density ${\bf f}_j$ on ${\mathbb A}$ by
\begin{equation}
\label{eq:marginal}
{\bf f}_j(z_j, \rho_j, t) = \langle {\mathcal F}(z_j, \hat z_j, \rho_j, \hat \rho_j ,t) , 1 \rangle_{\hat {\mathbb A}_j} , 
\end{equation}
where $\langle \cdot , \cdot \rangle_{\hat {\mathbb A}_j}$ denotes the duality between measures and functions of the variables $(\hat z_j, \hat \rho_j)$ in ${\mathbb A}^{N-1}$ (and ${\mathbb A}^{N-1}$ is denoted by $\hat {\mathbb A}_j$ when such a duality is considered). 
\label{def:marginal}
\end{definition}

We note that (\ref{eq:marginal}) is equivalent to saying that for any smooth function $\varphi_j(z_j,\rho_j)$ of the single variables $(z_j,\rho_j) \in {\mathbb A}$, we have
\begin{equation}
\label{eq:marginal_2}
\langle {\mathcal F}(z_j, \hat z_j, \rho_j, \hat \rho_j ,t) , \varphi_j (z_j, \rho_j)\rangle_{{\mathbb A}^N} = \langle {\bf f}_j(z_j, \rho_j, t) , \varphi_j(z_j, \rho_j) \rangle_{\mathbb A} . 
\end{equation}
To get an equation for ${\bf f}_j$ at the kinetic time-scale, we use the master equation in weak form (\ref{eq:master9}) with a test function $\varphi_j(z_j,\rho_j)$ of the single variables $(z_j,\rho_j) \in {\mathbb A}$. The resulting equation is given in section \ref{subsec:single_marginal}. It is not a closed equation because its coefficients depend on the full joint pdf ${\mathcal F}$. 

In order to obtain a closed system of equations, we make the assumption of propagation of chaos. Here, in the perspective of letting $N \to \infty$, we introduce a spatial variable $x_j = j/N$ and the cell-size $h = \frac{1}{N}$. We write ${\bf f}_j (z, \rho, t) = {\bf f}^h(x_j, z, \rho, t)$, where $(z,\rho) \in {\mathbb A}$. With these notations, the assumption of propagation of chaos reads:

\begin{assumption}
We assume that the joint pdf ${\mathcal F}(\vec z, \vec \rho, t)$ is written as:
\begin{eqnarray}
&&\hspace{-1.3cm}
{\mathcal F}(\vec z, \vec \rho, t) = \prod_{j=1}^N {\bf f}^h(x_j, z_j, \rho_j, t).
\label{eq:chaos}
\end{eqnarray}
\label{ass:chaos}
\end{assumption}

This assumption states that the cell-states and densities at different points are statistically independent. As a result, we obtain a closed kinetic equation for the one-particle marginal ${\bf f}^h(x_j, z_j, \rho_j, t)$ for a fixed number of cells $N$ in section \ref{subsec:chaos}. The next step is to make the number of cells $N \to \infty$ or equivalently, the cell-spacing $h = \frac{1}{N} \to 0$. For this purpose, we make the following mean-field assumption for the rates:

\begin{assumption}
We assume that as $h \to 0$ (or $N\to \infty$), and for any fixed $x$ and any subsequence $x_j=\frac{j}{N}$ such that $x_j \to x$, the following limit $\bar \gamma^h (x_j, z, \rho, t) \to \bar \gamma (x, z, \rho, t)$ exists, where 
\begin{eqnarray}
&&\hspace{-1cm}
\bar \gamma^h (x_j, z_j, \rho_j, t) := \big\langle \prod_{i \in \{1, \ldots , N\}, i \not = j } {\bf f}^h(x_i, z_i, \rho_i, t) , \gamma_j \, \, \big\rangle_{\hat {\mathbb A}_j} . 
\label{eq:effective_rate_chaos}
\end{eqnarray}
\label{ass:mean_field}
\end{assumption}

With these assumptions, we can first derive equations for the moments of the one-particle marginal in section \ref{subsec:largeN}, and then prove the convergence of the one-particle marginal distribution to a Dirac delta modeling a monokinetic distribution function in section \ref{subsec:dirac}. The final result is stated below: 

\begin{theorem}
We consider the one-particle marginal distribution $\tilde {\bf f}^h (x,z,\rho,t) = {\bf f}^h \big(x,z,\rho,$ $\frac{t}{h}\big) $ and let $h \to 0$. We assume that $\tilde {\bf f}^h \to \tilde{\bf f}$ where $\tilde {\bf f}$ is a measure of $(x,z,\rho,t)$ and that the convergence is as smooth as needed. We also assume the propagation of chaos assumption (Assumption \ref{ass:chaos}) and the mean-field limit assumption for the rates (Assumption \ref{ass:mean_field}). Then, formally, we have
\begin{eqnarray*}
&&\hspace{-1cm}
\tilde {\bf f} (x,\pm1,\rho,t) = p_\pm(x,t) \, \delta (\rho - \bar \rho(x,t)), 
\end{eqnarray*}
where $\bar \rho(x,t)$ and $p_\pm(x,t)$ satisfy the following system: 
\begin{eqnarray}
\partial_t \bar \rho + \partial_x ( \bar \rho \, u) &=& 0, 
\label{eq:rho_final}  \\
\partial_t u &=& \gamma_t ( u_{\mbox{\scriptsize coll}} - u ) , 
\label{eq:u_final}  
\end{eqnarray}
with 
\begin{eqnarray}
&&\hspace{-1cm}
\gamma_t = \tilde \gamma_- + \tilde \gamma_+, \qquad u_{\mbox{\scriptsize coll}} = \frac{\tilde \gamma_- - \tilde \gamma_+}{\tilde \gamma_- + \tilde \gamma_+}  ,
\label{eq:bar_gamma_final}  \\
&&\hspace{-1cm}
\tilde \gamma_\pm (x,t)= \bar \gamma (x, \pm 1, \bar \rho(x,t), t), 
\label{eq:bar_gamma_final_2} 
\end{eqnarray}
and with $\bar \gamma$ given by Assumption  \ref{ass:mean_field}, i.e. 
\begin{eqnarray}
&&\hspace{-1cm}
\bar \gamma^h (x_j, z, \rho, t) \to \bar \gamma (x, z, \rho, t) \quad \mbox{ as } \quad h \to 0, \nonumber \\
&&\hspace{-1cm}
\bar \gamma^h (x_j, z_j, \rho_j, t) := \big\langle \prod_{i \in \{1, \ldots , N\}, i \not = j } 
p_{z_i}(x_i,t) \, \delta (\rho_i - \bar \rho(x_i,t)) , \gamma_j \, \, \big\rangle_{\hat {\mathbb A}_j}. 
\label{eq:bar_gamma_final_3} 
\end{eqnarray}
Additionally, we have 
\begin{eqnarray}
&&\hspace{-1cm}
p_+ = \frac{1+u}{2}, \qquad p_- = \frac{1-u}{2}. 
\label{eq:mom_limits_final}  
\end{eqnarray}
\label{thm:hydro_limit}
\end{theorem}

The time rescaling (i.e. $t$ replaced by $t/h$ in the $1$-cell pdf) is needed to find the correct time-scale over which the pdf relaxes to an equilibrium. This time-scale is called the hydrodynamic time-scale, because it gives rise to the hydrodynamic model (\ref{eq:rho_final}), (\ref{eq:u_final}) (see comment below). It is a longer time-scale than the kinetic time-scale considered so far. This is because this relaxation is very slow and requires much longer time units to be observable. This hydrodynamic rescaling is classical in kinetic theory (see e.g. the review~\cite{Degond_Birkhauser03}).

Theorem \ref{thm:hydro_limit} states that in the limit $h \to 0$, the $1$-cell pdf $\tilde {\bf f}^h (x,z,\rho,t)$ observed at the hydrodynamic time-scale converges to a deterministic pdf in the density variable $\rho$, i.e. a Dirac delta located at the mean density $\bar \rho$. Both values of the pdf for the cell states $+1$ and $-1$ are proportional to the same Dirac delta, with proportionality coefficients $p_\pm$ meaning that among the $\bar \rho(x,t) \, dx$ particles located in the neighborhood $dx$ of position $x$ at time $t$, a proportion $p_+(x,t)$ (resp. $p_-(x,t)$) corresponds to right-going (resp. left-going) pedestrians (with $p_+(x,t)+p_-(x,t)=1$). Both the mean density $\bar \rho$ and the proportions $p_\pm$ depend of $(x,t)$. Their evolution is described by System (\ref{eq:rho_final}), (\ref{eq:u_final}). The mean velocity $u$ is given by (\ref{eq:mom_limits_final}) which shows that it is proportional to the imbalance between the right and left going pedestrians $u = p_+ - p_-$. 

Eq. (\ref{eq:rho_final}) is a classical continuity equation. It expresses that the total mass $M_{[a,b]}(t)$ contained in the interval $[a,b]$ at time (t) and given by $M_{[a,b]}(t)= \int_a^b \rho(x,t) \, dx$ evolves due to particles leaving or entering $[a,b]$ through its boundaries. Indeed, integrating (\ref{eq:rho_final}) with respect to $x \in [a,b]$, we get that 
$$ \frac{d}{dt} M_{[a,b]}(t) = (\rho u) (a,t) - (\rho u) (b,t). $$
The quantities $(\rho u) (a,t)$ and $(\rho u) (b,t)$ are the particles fluxes respectively through $a$ and $b$. These particle fluxes (counted positive if they are directed in the positive $x$ direction) contribute to an increase of the mass at $a$ and a decrease of the mass at $b$. Therefore,~(\ref{eq:rho_final}) describes a simple particle budget. 

By contrast, Eq. (\ref{eq:u_final}) is a simple ordinary differential equation describing the relaxation of the local velocity $u(x,t)$ to a velocity $u_{\mbox{\scriptsize coll}}(x,t)$ expressing a collective consensus. We will refer to this velocity as the collective consensus velocity. It depends on the state of the CA in a possibly large neighborhood of $x$ at time $t$. It is computed through (\ref{eq:bar_gamma_final}) in terms ot the switching rates of the cell corresponding to point $x$. More preciserly,  $u_{\mbox{\scriptsize coll}}(x,t)$ depends on the normalized difference between the switching rates for switching from state $-1$ to state $+1$ and for switching from state $+1$ to state $-1$. Indeed, this difference is the phenomenon producing a non-zero collective consensus velocity. There might be multiple solutions of the equation $u= u_{\mbox{\scriptsize coll}}$. These multiple solutions are associated to collective decision makings about the direction of the motion which can be according to the state of the CA, either left-going or right-going. In general, the actual velocity $u$ is different from the collective consensus velocity $u_{\mbox{\scriptsize coll}}$ and Eq. (\ref{eq:u_final}) states that $u$ relaxes to $u_{\mbox{\scriptsize coll}}$ at rate $\gamma_t$ equal to the sum of the switching rates. We will provide examples of these features in the next section. The fact that there is no spatial transport in (\ref{eq:u_final}) results from the instantaneous evaluation of the switching rates within the original CA. More sophisticated CA may result in the restoration of spatial transport in (\ref{eq:u_final}). Such dynamics will be studied in future work. 

The following sections are devoted to the proof of this theorem.

\subsection{Equation for the single-particle marginal distribution}
\label{subsec:single_marginal}

We remind that, in order to get an equation for ${\bf f}_j$ at the kinetic time-scale, we use the master equation in weak form (\ref{eq:master9}) with a test function $\varphi_j(z_j,\rho_j)$ of the single variables $(z_j,\rho_j) \in {\mathbb A}$. We have the following proposition, the proof of which is immediate and left to the reader:

\begin{proposition}
Define:
\begin{eqnarray}
\overline{\Delta \Psi_j}(t) \, {\bf f}_j(t) &:=& \langle {\mathcal F}(t) ,  \Psi_{j+\frac{1}{2}} - \Psi_{j-\frac{1}{2}} \rangle_{\hat {\mathbb A}_j} , \label{eq:effective_flux}\\
\bar \gamma^h_j(t) \,  {\bf f}_j (t) &:=& \langle {\mathcal F}(t) , \gamma_j \rangle_{\hat {\mathbb A}_j} . \label{eq:effective_rate}
\end{eqnarray}
The functions $\overline{\Delta \Psi_j}(t)$ and $\bar \gamma^h_j(t)$ are functions of $(z_j, \rho_j)$ only. Then, the equation for the marginal ${\bf f}_j$ is written in weak form: 
\begin{eqnarray}
&&\hspace{-1cm}
\langle \frac{\partial {\bf f}_j}{\partial t}, \varphi_j \rangle_{{\mathbb A}} 
=  - \langle {\bf f}_j , \overline{\Delta \Psi_j} \,\partial_{\rho_j} \varphi_j  \, \rangle_{{\mathbb A}} 
+   \frac{1}{N} \, \langle \,  {\bf f}_j , \bar \gamma^h_j \,  \{ \varphi_j(-z_j, \rho_j) - \varphi_j(z_j, \rho_j) \} \, \big\rangle_{{\mathbb A}} , 
\label{eq:marginal1}
\end{eqnarray}
and in strong form
\begin{eqnarray}
&&\hspace{-1cm}
\Big( \frac{\partial}{\partial t}  {\bf f}_j 
-  \partial_{\rho_j} \big( \overline{\Delta \Psi_j} \, {\bf f}_j  \big) \Big) (x_j, z_j, \rho_j, t)
  \nonumber \\
&&\hspace{0cm}
= \frac{1}{N} \, \big( \bar \gamma^h_j (x_j, -z_j, \rho_j,t) {\bf f}_j (x_j, -z_j, \rho_j, t) 
- \bar \gamma^h_j(x_j, z_j, \rho_j, t) {\bf f}_j (x_j, z_j, \rho_j, t) \big)  .
\label{eq:marginal2}
\end{eqnarray}
\label{prop:marginal_eq}
\end{proposition}

We introduce the following definition of moments and velocity:

\begin{definition} 
The probabilities of having right-going (respectively left-going) particles at $(x,t)$ is denoted by $p_+(x,t)$ (resp. $p_-(x,t)$). The average right-going (respectively left-going) particle density at $(x,t)$ is denoted by $\bar \rho_+(x,t)$ (resp. $\bar \rho_-(x,t)$). They are defined by:
\begin{eqnarray}
&&\hspace{-1.3cm}
p_{\pm}(x,t) = \int_0^\infty {\bf f}(x, \pm 1, \rho, t) \, d \rho, \quad \bar \rho_{\pm} (x,t) = \int_0^\infty {\bf f}(x, \pm 1, \rho, t) \, \rho \, d \rho. 
\label{eq:moments}
\end{eqnarray}
The average velocity of the particles at $(x,t)$ is defined by:
\begin{eqnarray}
&&\hspace{-1.3cm}
u(x,t) = (p_+ - p_-)(x,t). 
\label{eq:velocity}
\end{eqnarray}
We note that $p_\pm$ and $\bar \rho_\pm$ are non-negative quantities and that $p_+ + p_- = 1$. We define $\bar \rho = \bar \rho_+ + \bar \rho_-$ the total particle density at $(x,t)$. 
\label{def:moments}
\end{definition}

In the following section, we use the propagation of chaos assumption to close the kinetic equation (\ref{eq:marginal2}) for the one-particle marginal distribution.

\subsection{Propagation of chaos assumption and closed kinetic equation for the one-particle marginal distribution}
\label{subsec:chaos}

We now make the propagation of chaos assumption (Assumption \ref{ass:chaos}). With this assumption we can simplify the expressions of the flux (\ref{eq:effective_flux}). We have the following: 

\begin{lemma}
Under the chaos assumption (Assumption \ref{ass:chaos}), the flux (\ref{eq:effective_flux}) is given by:
\begin{eqnarray}
&&\hspace{-1cm}
\overline{\Delta \Psi_j}(t) = \overline{\Delta \Psi_j}(\rho_j,t) = \rho_j - \bar \rho_-(x_j + h, t) -  \bar \rho_+(x_j - h, t), 
\label{eq:eff_flux_chaos} 
\end{eqnarray}
\label{lem:flux_rates}
\end{lemma}

\noindent
{\bf Proof:} By direct computation from (\ref{eq:flux}), we have 
$$ \Psi_{j+\frac{1}{2}} - \Psi_{j-\frac{1}{2}} = \rho_j + \rho_{j+1} \min \{ z_{j+1}, 0 \} - \rho_{j-1} \max\{ z_{j-1}, 0 \} . $$
So, now, 
\begin{eqnarray}
&&\hspace{-1cm}
\big\langle {\mathcal F}(t) , \Psi_{j+\frac{1}{2}} - \Psi_{j-\frac{1}{2}} \big\rangle_{\hat {\mathbb A}_j} = {\bf f}(x_{j}, z_{j}, \rho_{j}, t) \big\langle  {\bf f}(x_{j-1}, z_{j-1}, \rho_{j-1}, t) {\bf f}(x_{j+1}, z_{j+1}, \rho_{j+1}, t), \nonumber\\
&&\hspace{3cm}
 \rho_j + \rho_{j+1} \min \{ z_{j+1}, 0 \} - \rho_{j-1} \max\{ z_{j-1}, 0 \}  \big\rangle_{{\mathbb A}_{j-1} \otimes {\mathbb A}_{j+1}} , \label{eq:eff_flux_2}
\end{eqnarray}
where $\langle \cdot , \cdot \rangle_{{\mathbb A}_{j-1} \otimes {\mathbb A}_{j+1}}$ denotes the duality between measures and functions on ${\mathbb A}^2$ with respect to the variables $(z_{j-1}, \rho_{j-1}, z_{j+1}, \rho_{j+1})$. Then, using the definitions of the moments (\ref{eq:moments}), the evaluation of the right-hand side of (\ref{eq:eff_flux_2}) leads to (\ref{eq:eff_flux_chaos}). \endproof

As in the previous section we  index the one-particle marginal distribution by $h = \frac{1}{N}$ and denote it by ${\bf f}^h$ and similarly we denote by  $ \overline{\Delta \Psi}^h(x_j,z,\rho,t)= \overline{\Delta \Psi}_j(\rho,t)$. With Lemma \ref{lem:flux_rates}, we can get a closed equation for ${\bf f}^h$ . More precisely, we have the following:
 
\begin{proposition}
Under the propagation of chaos assumption (Assumption \ref{ass:chaos}), the single-particle marginal distribution function ${\bf f}^h$ satisfies the closed kinetic equation: 
\begin{eqnarray}
&&\hspace{-1.3cm}
\Big( \frac{\partial}{\partial t}  {\bf f}^h 
-  \partial_{\rho} \big( \overline{\Delta \Psi^h } \, {\bf f}^h  \big) \Big) (x_j, z, \rho, t) \nonumber \\
&&\hspace{1.3cm}
=  h \,  \big( \bar \gamma^h(x_j, -z, \rho,t) {\bf f}^h (x_j, -z, \rho, t) 
- \bar \gamma^h(x_j, z, \rho, t) {\bf f}^h (x_j, z, \rho, t) \big)  .
\label{eq:marginal3}
\end{eqnarray}
with rates given by (\ref{eq:effective_rate_chaos}). 
\label{prop:kinetic_closed}
\end{proposition}

Now we make a change of time scale to the macroscopic time scale. We let $t' = h t$. The rationale for this change is that both $\overline{\Delta  \Psi}^h$ and the right-hand side of (\ref{eq:marginal3}) formally tend to zero as $h \to 0$. In order to recover a meaningful dynamics for the one-particle marginal, we have to observe it on a time interval of length $1/h$. Performing this change of variable in (\ref{eq:marginal3}) and dropping the primes for simplicity, we are led to the following problem: 

\begin{eqnarray}
&&\hspace{-1.3cm}
\Big( \frac{\partial}{\partial t}  {\bf f}^h 
-  \partial_{\rho} \big( \frac{1}{h} \, \overline{\Delta \Psi}^h \, {\bf f}^h  \big) \Big) (x_j, z, \rho, t) \nonumber \\
&&\hspace{1.3cm}
=  \bar \gamma^h(x_j, -z, \rho,t) {\bf f}^h (x_j, -z, \rho, t) 
- \bar \gamma^h(x_j, z, \rho, t) {\bf f}^h (x_j, z, \rho, t)  .
\label{eq:marginal3.1}
\end{eqnarray}

In the next section, we investigate the $h \to 0$ limit. A key assumption will be that the rates converge to their mean-field limit, as stated in Assumption \ref{ass:mean_field}.

\subsection{Large cell-number mean-field limit and macroscopic moments}
\label{subsec:largeN}

In this section, we make the formal limit of a large number of cells $N \to \infty$ or $h \to 0$. 
We assume that ${\bf f}^h \to {\bf f}$ where ${\bf f}$ is a measure of $(x,z,\rho,t)$ and that the convergence is as smooth as needed. The goal of this section is to compute the dynamics of ${\bf f}$. For this purpose, we need Assumption \ref{ass:mean_field} which assumes that the rates converge to their mean-field limit. This assumption will be shown for some example in section \ref{sec:examples} below. We first consider the equations for the total density $\bar \rho$ given by (\ref{eq:moments}) and the mean velocity $u(x,t)$ given by (\ref{eq:velocity}). We have the:

\begin{lemma}
When $h \to 0$, we formally have $\bar \rho^h \to \bar \rho$ and $u^h \to u$ where $\bar \rho$ and $u$ satisfy:
\begin{eqnarray}
\partial_t \bar \rho + \partial_x ( \bar \rho \, u) &=& 0, 
\label{eq:rho_largeN}  \\
\partial_t u &=& \gamma_t ( u_{\mbox{\scriptsize coll}} - u ) , 
\label{eq:u_largeN}  
\end{eqnarray}
with 
\begin{eqnarray}
&&\hspace{-1cm}
\gamma_t = \gamma^{(0)}_- + \gamma^{(0)}_+, \qquad u_{\mbox{\scriptsize coll}} = \frac{\gamma^{(0)}_- - \gamma^{(0)}_+}{\gamma^{(0)}_- + \gamma^{(0)}_+}  ,\label{eq:bar_gamma}  
\end{eqnarray}
defining, 
\begin{eqnarray}
&&\hspace{-1cm}
\gamma^{(k)}_\pm (x,t)   = \frac{\int_0^\infty \bar \gamma(x, \pm 1,\rho,t) {\bf f} (x, \pm 1, \rho, t) \, \rho^k \, d \rho}{\int_0^\infty {\bf f} (x, \pm 1, \rho, t) \, \rho^k \, d \rho}.
\label{eq:rates_gamma_k}
\end{eqnarray}
We note that the denominator of the expression (\ref{eq:rates_gamma_k}) with $k=0$ of $\gamma^{(0)}_\pm$ is $p_\pm$ and that: 
\begin{eqnarray}
&&\hspace{-1cm}
p_+ = \frac{1+u}{2}, \qquad p_- = \frac{1-u}{2}. 
\label{eq:mom_limits}  
\end{eqnarray}
\label{lem:limit_largeN}
\end{lemma}

\noindent
{\bf Proof:} By Taylor expansion and since $\bar{\rho}=\bar{\rho}_+ + \bar{\rho}_-$, we have: 
\begin{eqnarray}
&&\hspace{-1cm}
\frac{1}{h} \overline{\Delta \Psi_j^h}(\rho_j,t) =\frac{1}{h} \big( \rho_j - \bar \rho(x_j,t) \big) + \partial_x ( \bar \rho_+ - \bar \rho_-)(x_j,t) + o(h), 
\label{eq:eff_flux_taylor} 
\end{eqnarray}
Inserting this expansion into (\ref{eq:marginal2}) and using the mean field assumption for rates (Assumption \ref{ass:mean_field}), we have 
\begin{eqnarray}
&&\hspace{-1.3cm}
\Big( \frac{\partial}{\partial t}  {\bf f}^h 
-  \partial_{\rho} \big( \partial_x (\bar \rho_+ - \bar \rho_-)  \, {\bf f}^h  \big) \Big) (x, z, \rho, t) = \frac{1}{h} \partial_{\rho} \Big( \big( \rho - \bar \rho(x,t) \big)  {\bf f}^h \Big)(x, z, \rho, t) \nonumber \\
&&\hspace{1.3cm}
+  \bar \gamma(x, -z,\rho,t) {\bf f}^h (x, -z, \rho, t) 
- \bar \gamma(x, z, \rho, t) {\bf f}^h (x, z, \rho, t) + o(h)  .
\label{eq:marginal4}
\end{eqnarray}
Now, multiplying (\ref{eq:marginal4}) by $\rho$ and integrating with respect to $\rho \in {\mathbb R}_+$ fixing $z$ to the values $z=+1$ and $z=-1$ successively, we get: 
\begin{eqnarray*}
&&\hspace{-1.3cm}
\Big( \frac{\partial}{\partial t}  \bar \rho_+^h 
+ p_+^h \big( \partial_x (\bar \rho_+ - \bar \rho_-)  \big) \Big) (x, t) = - \frac{1}{h} \big( \bar \rho_+^h - p_+^h \bar \rho \big)  (x, t) \nonumber \\
&&\hspace{6.3cm}
+  \big( \gamma^{(1)}_- \bar \rho_-^h  
-  \gamma^{(1)}_+ \bar \rho_+^h \big)(x, t) + o(h) , \\
&&\hspace{-1.3cm}
\Big( \frac{\partial}{\partial t}  \bar \rho_-^h 
+ p_-^h \big( \partial_x (\bar \rho_+ - \bar \rho_-)  \big) \Big) (x, t) = - \frac{1}{h} \big( \bar \rho_-^h - p_-^h \bar \rho \big)  (x, t) \nonumber \\
&&\hspace{6.3cm}
+  \big( \gamma^{(1)}_+ \bar \rho_+^h - \gamma^{(1)}_- \bar \rho_-^h \big)(x, t) + o(h)  .
\end{eqnarray*}
Adding and subtracting these two equations, we get:
\begin{eqnarray}
&&\hspace{-1.3cm}
\Big( \frac{\partial}{\partial t}  \bar \rho^h 
+  \partial_x (\bar \rho_+ - \bar \rho_-)  \Big) (x, t) = o(h)  , 
\label{eq:moments_sum} \\
&&\hspace{-1.3cm}
\Big( \frac{\partial}{\partial t}  ( \bar \rho_+^h - \bar \rho_-^h )
+ ( p_+^h - p_-^h)  \big( \partial_x (\bar \rho_+ - \bar \rho_-)  \big) \Big) (x, t) = - \frac{1}{h} \big( \bar \rho_+^h - \bar \rho_-^h - ( p_+^h - p_-^h) \bar \rho \big)  (x, t) \nonumber \\
&&\hspace{6.3cm}
+  2 \big( \gamma^{(1)}_- \bar \rho_-^h - \bar \gamma^{(1)}_+ \bar \rho_+^h \big)(x, t) + o(h)  ,
\label{eq:moments_diff} 
\end{eqnarray}
Now, letting $h \to 0$ in (\ref{eq:moments_diff}), leads to 
\begin{eqnarray*}
\bar \rho_+ - \bar \rho_- = (p_+ - p_-) \bar \rho = u \bar \rho
\end{eqnarray*}
and inserting it into (\ref{eq:moments_sum}) leads to the conservation equation (\ref{eq:rho_largeN}).

Now, multiplying (\ref{eq:marginal4}) by $1$, integrating with respect to $\rho \in {\mathbb R}_+$ fixing $z$ to the value $z=1$ and letting $h \to 0$, we get: 
\begin{eqnarray*}
&&\hspace{-1.3cm}
\frac{\partial}{\partial t}  p_+  (x, t) =  \big( \gamma^{(0)}_- p_-  
- \gamma^{(0)}_+ p_+ \big)(x, t) + o(h)  . 
\end{eqnarray*}
Now, using that $p_- = 1 - p_+$, simple algebraic manipulations lead to (\ref{eq:u_largeN}). Finally, eqs.  (\ref{eq:mom_limits}) are obvious from what precedes. This ends the proof. \endproof

So far, system (\ref{eq:rho_largeN}), (\ref{eq:u_largeN}) is not closed because we are lacking a simple expression of $\gamma^{(0)}_\pm$ in terms of $\bar \rho$ and $u$. In the next section, we provide such a closure relation by taking the limit $h \to 0$ in the kinetic equation (\ref{eq:marginal4}).

\subsection{Local Equilibrium closure and macroscopic model}
\label{subsec:dirac}

We now consider (\ref{eq:marginal4}) and let  $h \to 0$ in it. We have the following: 

\begin{lemma}
Let ${\bf f} = \lim_{h \to 0} {\bf f}^h$. Then, ${\bf f}$ is written
\begin{eqnarray}
&&\hspace{-1cm}
{\bf f} (x,1,\rho,t) = p_+(x,t) \, \delta (\rho - \bar \rho(x,t)), \quad {\bf f} (x,-1,\rho,t) = p_-(x,t) \, \delta (\rho - \bar \rho(x,t)), 
\label{eq:dirac_closure}  
\end{eqnarray}
where $p_\pm$ and $\bar \rho$ are the moments defined at Definition \ref{def:moments}. This leads to the following expression of $\gamma^{(0)}_\pm$:  
\begin{eqnarray}
&&\hspace{-1cm}
\gamma^{(0)}_\pm (x,t)   = \frac{1}{p_\pm(x,t)} \bar \gamma(x, \pm 1, \bar \rho(x,t) ,t).
\label{eq:rates_gamma_0}
\end{eqnarray}
\label{lem:Dirac_closure}
\end{lemma}

\noindent
{\bf Proof:} Taking $h \to 0$ in (\ref{eq:marginal4}), we are led to the fact that ${\bf f}$ satisfies:
$$ \partial_{\rho} \Big( \big( \rho - \bar \rho \big) \, {\bf f} \Big) = 0 , $$
which implies, since ${\bf f}$ must be a positive measure, that
\begin{equation}
{\bf f} (x,z,\rho,t) = p(x,z,t) \delta (\rho - \bar \rho (x,t)), 
\label{eq:equi}
\end{equation}
with a convenient $p(x,z,t)$. Additionally, if we focus on the leading order term, we can consider the simplified problem:  
\begin{eqnarray*}
&&\hspace{-1.3cm}
\frac{\partial}{\partial t}  {\bf f}^h  - \frac{1}{h} \partial_{\rho} \Big( \big( \rho - \bar \rho \big) \, {\bf f}^h \Big) =0, 
\end{eqnarray*}
This is a first order partial differential equation which can be solved by characteristics. We denote by $\rho(t)$ an arbitrary characteristics. It is obtained by solving the equation. 
$$ \dot \rho (t) = - \frac{1}{h}  \big( \rho - \bar \rho \big), $$
Its solution converges in exponential time with time-scale $O(h)$ towards the fixed point $\bar \rho$. Therefore, ${\bf f}^h$ itself converges in exponential time towards a distribution of the form (\ref{eq:equi}). Now, by taking the moments of (\ref{eq:equi}), we realize that the LE has necessarily the form (\ref{eq:dirac_closure}). Inserting this expression into (\ref{eq:rates_gamma_k}) (with $k=0$) leads to (\ref{eq:rates_gamma_0}). \endproof

This lemma completes the proof of Theorem \ref{thm:hydro_limit}.

\setcounter{equation}{0}
\section{Example: a model for pedestrian flow in corridors}
\label{sec:examples}

Here, we are interested in pedestrian dynamics within a corridor. We assume that the corridor is decomposed into small stretches (the cells) and that within a given cell, the flow of pedestrians is either left or right-going. The orientation of the flow in this cell is described by the variable $z_j^n$ ($z_j^n = +1$ if the flow is right-going, and $z_j^n = -1$ if the flow is left-going). The orientation of the flow is controlled by which of these two flows is the largest. If the right-going flux is larger than the left-going one, then the probability that the state of the cell is given by $z=+1$ increases, i.e. if the state is already $z=+1$, it will have a larger probability to stay at this value, while if the state is originally $z=-1$, the probability for a state-change to the value $z=+1$ increases.

To model this rule, we assume that  the rate of change for cell $j$ at time $t_n$ can be given by 
\begin{equation}
\gamma_j^n := \gamma_j( z_j^n, {\hat{z}}^n_j, \rho^n) = \gamma_0 + b | z_j^n -  \langle z \rangle_j^n |^\alpha
\label{eq:rate}
\end{equation}
where  
\begin{equation}
\langle z \rangle_j^n = \frac{ \frac{1}N \sum\limits_{i=1}^N z_i^n \,  w\big( \frac{i-j}{N} \big) \, \pi(\rho_i^n) }{ \frac{1}N \sum\limits_{i=1}^N w \big( \frac{i-j}{N} \big) \, \pi(\rho_i^n) }.
\label{eq:avg}
\end{equation}
The coefficients $\gamma_0$ and $b$ are supposed to be non--negative (and might as well depend on $j$ and $n$).
We assume that $\alpha \geq 0$ and that the density-sensing function $\pi$ is supposed to be monotone increasing with $\pi(0)=0.$ The weight $w:[0,1]\to{\mathbb R}_+$ is a smooth function. We note that, because $z_j^n = \pm 1$, $-1 \leq \langle z \rangle_j^n  \leq 1$.

The rationale for (\ref{eq:rate}), (\ref{eq:avg}) is as follows. The quantity $\langle z \rangle_j^n$ describes the state of the given cell and the neighbouring ones, defined by those which are in the support of the function $w$. This average weights the cells with a large density more strongly than those with a low density thanks to the density-sensing function $\pi$. Now, the probability for a cell-state change is decreased if the actual state variable $z_j^n$ is close to the average $\langle z \rangle_j^n$, while it incerases if the distance to the average $\langle z \rangle_j^n$ increases. This increase is linear if $\alpha = 1$ and super-linear if $\alpha >1$. A super-linear increase triggers self-organization as we will see below, while a linear increase does not. In addition to cell-state changes due to pedestrians interaction as just described, we add a certain level of fluctuations described by a constant rate of cell-state changes equal to $\gamma_0$. Many modeling choices for the kernel $w$ can be envisioned. For instance, a symmetric weighting function $w$ parameterized by a sensing radius $r>0$ of the form:
\begin{eqnarray}
\label{eq:w} 
w(x)=w_r(x)= \frac{1}{\sqrt{\pi} \, r} \, \exp( - \frac{x^2}{r^2} ), 
\end{eqnarray}
can be chosen. 
Here, $r$ is kept fixed and $O(1)$. 

In this example, we verify the mean-field assumption for the rates (Assumption \ref{ass:mean_field}), as shown in the following:

\begin{lemma}
In the limit $h \to 0$, we formally have $\bar \gamma^h (x_j, z, \rho, t) \to \bar \gamma (x_j, z, t)$ with  
\begin{eqnarray}
&&\hspace{-1cm}
\bar \gamma (x,z,t)  =  \gamma_0 + b \, \left| \,  z -  \frac{   \sum_{\zeta=\pm 1} \int_{(y,\xi) \in [0,1] \times {\mathbb R}_+} \zeta \, w(y-x) \, \pi (\xi) \, {\bf f}(y, \zeta, \xi, t) \, d \xi \, dy }{\sum_{\zeta=\pm 1} \int_{(y,\xi) \in [0,1] \times {\mathbb R}_+} w(y-x) \, \pi (\xi) \, {\bf f}(y, \zeta, \xi, t) \, d \xi \, dy } \, \right|^\alpha. \nonumber \\
&&\hspace{-1cm}
\label{eq:eff_rate_largeN} 
\end{eqnarray}
In particular, $\bar \gamma (x,z,t)$ does not depend on $\rho$. 
\label{lem:eff_rate}
\end{lemma}

\noindent
{\bf Proof:} Formula (\ref{eq:effective_rate_chaos}) can be written as
\begin{eqnarray*}
&&\hspace{-1cm}
\bar \gamma^h(x_j,z_j,\rho_j,t)  =  \\
&&\hspace{-0.5cm}
= \sum_{\hat z_j \in \{-1,1\}^{N-1}} \int_{(\hat x_j, \hat \rho_j) \in ([0,1] \times {\mathbb R}_+)^{N-1}} 
\Big( \gamma_0 + b \, \Big| \,  z_j -  \frac{  \frac{1}{N} \sum_{i=1}^{N}  z_i \, w\big( x_i - x_j \big) \, \pi(\rho_i)}{\frac{1}{N}\sum_{i=1}^{N}  w \big(  x_i - x_j \big) \, \pi(\rho_i)  }  \, \Big|^\alpha \Big) \times \\
&&\hspace{8cm}
 \prod_{k=1, k \not = j}^N  {\bf f}^h(x_k,z_k, \rho_k,t)  \, \, d \hat x_j \, d \hat \rho_j 
\end{eqnarray*}
The numerator and denominator of the fraction inside the integral are mean values of the functions 
$(y,\zeta, \xi) \in [0,1] \times  \{-1,1\} \times {\mathbb R}_+ \to \zeta \, w(y-x) \, \pi (\xi)$ and 
$(y,\zeta, \xi) \to w(y-x) \, \pi (\xi)$ respectively, 
over $N-1$ independent identically distributed random variables $(x_i, z_i, \rho_i)$ drawn according to the probability distribution 
${\bf f} = \lim_{h\to 0} {\bf f}^h$. Therefore, for large $N$, they converge to the average value of these functions respectively, which make the numerator and denominator of (\ref{eq:eff_rate_largeN}). Then by formal manipulation, we deduce (\ref{eq:eff_rate_largeN}). The proof of this result, which requires the central limit theorem is outside the scope of this paper. \endproof

\begin{figure}[htbp]
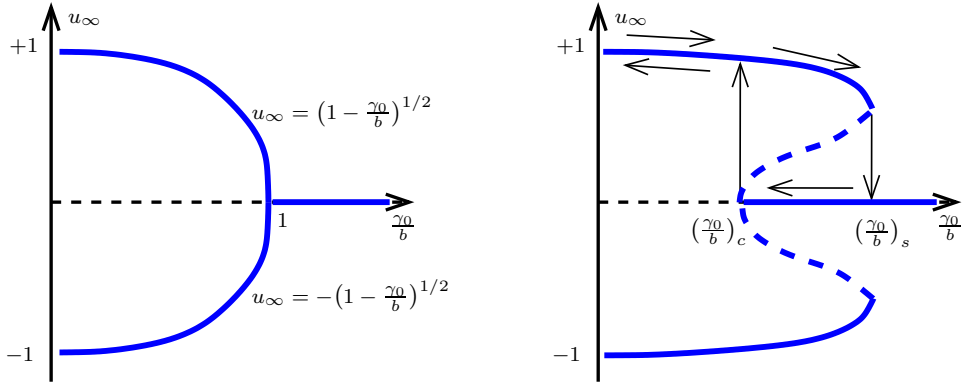

\begin{center}
\input{fork.pstex_t}
\hspace{1.5cm}
\input{hyst.pstex_t}
\caption{Left: pitchfork bifurcation diagram for the stationary equilibrium solution $u_\infty$ for $\alpha = 2$. Right: sub-critical pitchfork bifurcation diagram for the stationary equilibrium solution $u_\infty$ for $\alpha >6$. The arrows highlight the hysteresis loop. }
\label{fig:fork}
\end{center}
\end{figure}

Within this example, Theorem \ref{thm:hydro_limit} holds true with $u = p_+ - p_-$ and 
\begin{eqnarray}
&&\hspace{-1cm}
\tilde \gamma_+ (x,t)  =  \gamma_0 + b \, \big| \,  1 -  \langle u \rangle (x,t) \, \big|^\alpha , 
\qquad 
\tilde \gamma_- (x,t)  =  \gamma_0 + b \, \big| \,   - 1 -  \langle u \rangle (x,t) \, \big|^\alpha , 
\label{eq:gamma+-_delta} \\
&&\hspace{-1cm}
\langle u \rangle (x,t) = \frac{ \int_{y \in [0,1]} w(y-x) \, \pi (\bar \rho(y,t)) \, u(y,t) \, dy }{\int_{y \in [0,1]} w(y-x) \, \pi (\bar \rho(y,t))\, dy } . 
\label{eq:<u>_delta}  
\end{eqnarray}
In the case $\alpha = 2$, we note that $ \gamma_t = 2 ( \gamma_0 + b(1 + \langle u \rangle^2))$ and $u_{\mbox{\scriptsize coll}} = \frac{2 b \langle u \rangle}{\gamma_0 + b(1 + \langle u \rangle^2)}$. If we restrict ourselves to spatially homegeneous solutions, then $\bar \rho$ is uniform and constant and $\langle u \rangle = u $ only depends on time. Furthermore, $\langle u \rangle$ is independent of $\pi$. Then, inserting this into (\ref{eq:gamma+-_delta}) leads to the third order differential equation:
\begin{eqnarray}
&&\hspace{-1cm}
\partial_t u = 2 b u \big[ 1-\frac{\gamma_0}{b} - u^2 \big].
\label{eq:u_hom}  
\end{eqnarray}
The parameter $\gamma_0/b$, which describes the ratio of the noise to consensus force can be seen as a bifurcation parameter for this Ordinary Differential Equation. This ODE has a pitchfork bifurcation with critical point $\gamma_0/b = 1$. Indeed, the equilibrium solutions of this equation when $t \to \infty$ are $u_\infty=0$ or $u_\infty^2 =  \frac{b-\gamma_0}{b}$. Therefore, if $\gamma_0>b$, $u_\infty=0$ is the only stationary equilibrium and it can be seen that it is a stable one (the right-hand side of (\ref{eq:u_hom}) has opposite sign to $u$). By contrast if $\gamma_0<b$, two other stationary equilibria exist: $u_\infty =  \pm \frac{\sqrt{b-\gamma_0}}{\sqrt{b}}$. Then, it is readily seen by inspection of (\ref{eq:u_hom}) that the equilibrium $u_\infty=0$ is unstable while the two equilibria $u_\infty =  \pm \frac{\sqrt{b-\gamma_0}}{\sqrt{b}}$ are stable. In this case, the stable equilibrium describes the formation of a consensus about one direction of motion. This consensus is obeyed by all the more people that the random state-change frequency $\gamma_0$ is close to $0$. This analysis shows that there exists a phase transition from disordered to ordered motion when $b$ (which describes the consensus force) crosses $\gamma_0$. The bifurcation diagram is shown in Fig. \ref{fig:fork} (left). The upper half of the curve (which provides the order parameter $|u_\infty|$ versus the noise level $\gamma_0/b$) can be regarded as the standard phase-transition diagram. In this case, this diagram indicates a second-order (or continuous) phase-transition with critical exponent $1/2$. 

By contrast, in the case $\alpha = 1$, we find $\bar \gamma_t = 2 ( \gamma_0 + b)$ and $u_{\mbox{\scriptsize coll}} = \frac{b \langle u \rangle}{\gamma_0 + b}$. If a spatially homogeneous solution is sought, it is given by 
\begin{eqnarray}
&&\hspace{-1cm}
\partial_t u = - 2 \gamma_0 u .
\label{eq:u_hom_1}  
\end{eqnarray}
Then, the stationary equilibrium solution $u_\infty=0$ is the only solution. There is no possibility of ordered motion. The motion stays disordered whatever the value of the consensus force $b$ is. Therefore, an exponent strictly larger than $1$ is necessary for the appearance of consensus. 

In the case $\alpha \geq1$ and for spatially homogeneous problems, a general formulation of the equation for $u$ is available as follows: 
\begin{eqnarray}
&&\hspace{-1cm}
\partial_t u =  2 b u \, \Big[ - \frac{\gamma_0}{b} + (1-u^2) \, \frac{(1+u)^{\alpha-1} - (1-u)^{\alpha-1}}{2 u} \Big] .
\label{eq:u_hom_alpha}  
\end{eqnarray}
We recover (\ref{eq:u_hom_1}) and (\ref{eq:u_hom}) in the cases $\alpha = 1$ and $\alpha =2$ respectively. For integers $\alpha = 3$ up to $\alpha = 5$, the behavior is the same as for $\alpha = 2$ with the critical point becoming $(\gamma_0/b)_c = \alpha-1$. For integers $\alpha >6$, there is another critical point $(\gamma_0/b)_s > (\gamma_0/b)_c = \alpha -1$  and the bifurcation diagram shows a sub-critical pitchfork bifurcation as depicted in Fig. \ref{fig:fork} (right). Arrows in Fig. \ref{fig:fork} (right) indicate the existence of a hysteresis loop. As before, the upper half of this diagram provides the phase-transition diagram giving the order parameter $|u_\infty|$ as a function of the noise level $\gamma_0/b$. In this case, this diagram indicates a first-order (or discontinuous) phase-transition as shown by the occurrence of a jump at the value $(\gamma_0/b)_s$.

\setcounter{equation}{0}
\section{Networks}
\label{sec:network}

\subsection{Graph framework}
\label{sub:graph}

The goal of this section is to extend the previous CA and its associated mean-field and hydrodynamic limits to more general network topologies. We consider a network as a graph $(\mathcal{J},\mathcal{A})$ where $\mathcal{A}$ is the set of graph edges and $\mathcal{J}=\{1, \dots, J\}$ is the set of graph nodes.  We denote by $J = \mbox{Card}\, {\mathcal J}$. We define the adjacency matrix $(a_{jk})_{j,k \in {\mathcal J}}$, i.e., the matrix such  that $a_{jk} = a_{kj} = 1$ when node $j$ is connected to node  $k$  and $0$ otherwise. We assume that the nodes are not connected to themselves $a_{jj}=0.$ 
 We denote by $d_j$ the degree of node $j$, i.e. $d_j = \sum_{k \in {\mathcal J}} a_{jk}$. 
We assume that the graph is connected, i.e. for any pair of nodes $(j,k)$, with $k \not = j$, there exists a path withing the graph which connects $j$ and $k$. For each node $j$, we define the set ${\mathcal N}_j$ of nodes connected to it, i.e. 
$$ {\mathcal N}_j = \{ k \in {\mathcal J} \, | \, a_{jk} = 1 \}, $$
with $\mbox{Card} \, {\mathcal N}_j = d_j$. 

Each node $j \in {\mathcal J}$ contains the density $\rho_j \geq 0$ of the sweeping quantity. Indeed, this quantity is able to sweep from node $j$ to any other (directly) connected node $k$ (i.e. such that $a_{jk} = 1$). We assume that the whole quantity $\rho_j$ sweeps entirely to one of the neighboring nodes. Of course, a more complex model can be envisionned but we wish to keep the setting as simple as possible for this presentation. We denote by $\Psi^n_{jk}$ the outgoing flux from $j$ to $k$ at time $t^n$. Each node carries the index of the neighboring node to which it sweeps $z_{j} \in {\mathcal N}_j$. 
Then, the outgoing flow (counted algebraically) from $j$ to $k$, denoted by $\Psi^n_{jk}$ is given by 
\begin{equation}
\Psi^n_{jk} =  \rho_j^n \delta_{z_j \, k} - \rho_k^n \delta_{z_k \, j}, 
\label{eq:flux_network}
\end{equation}	
where $\delta_{ij}$ is the Kronecker index: $\delta_{ij} = 1$ if $i=j$ and $0$ otherwise. The convention is that the flux between $j$ and $k$ is positive when it is outgoing from $j$ and negative when it is incoming. With this convention, we have $\Psi^n_{kj}= - \Psi^n_{jk}$.

Now, node $j$ changes state according to a Poisson process with rate $\gamma_{j}^n$ depending on the states and densities of the nodes in the vicinity of $j$. Within a given time interval ${\Delta t}$ the probability to change the state of node $j$ is $1 - \exp \big( -\gamma_{j}^n \Delta t \big)$, and the change from state $z_j$ to any other state $z_j' \in {\mathcal N}_j \setminus \{z_j\}$ occurs with uniform probability. This means: 
\begin{equation}
z_{j}^{n+1 } = \left\{ \begin{array}{l} z_{j}^n \, \, \mbox{with probability} \, \, e^{-\gamma_j^n \Delta t}, \\
z_j' \in {\mathcal N}_j \setminus \{z_j\} \, \, \mbox{uniformly in} \, \,  {\mathcal N}_j \setminus \{ z_{j}^n \} \, \, \mbox{each with probability} \, \, \frac{1- e^{-\gamma_j^n \Delta t}}{d_j-1}. \end{array} \right.
\label{eq:jumpprocess_net}
\end{equation}
Given some initial data $z_{j}^0$ and $\rho_j^0$ for $j \in {\mathcal J}$, the discrete time update algorithm for $\rho_j^n$ is given at any discrete time index $n \in {\mathbb N}$ by: 
\begin{equation}\label{eq:micromodel_net} 
\rho_j^{n+1} = \rho_j^n - J \Delta t \sum_{k \in {\mathcal N}_j} \Psi_{jk}^n .
\end{equation} 

\begin{remark}
In the one-dimensional case of section \ref{sec:micro}, the vertices of the graph are the centers of the cells. 
\label{rem:link_to_1D}
\end{remark}

Now, we have the following proposition, which shows that the total number of particles is conserved:

\begin{proposition}
(i) The total number of particles is conserved, i.e. 
$$ \sum_{j\in{\mathcal J}} \rho_j^n = \sum_{j\in{\mathcal J}} \rho_j^0.$$ 
(ii) (Positivity preservation) Introduce $d = \max_{j \in {\mathcal J}} d_j$ the maximal degree of the nodes. Suppose that the CFL condition $J \Delta t \leq  \frac{1}{d}$ is satisfied. Then, we have
\begin{eqnarray}
&&\hspace{-1cm}
\rho_j^{n} \geq 0, \quad \forall j \in {\mathcal J} \quad \quad \Longrightarrow \quad \quad \rho_j^{n+1} \geq 0, \quad \forall j \in {\mathcal J}. 
\label{eq:max_princ}
\end{eqnarray}
\label{prop:max_princ}
\end{proposition}

\medskip
\noindent
{\bf Proof.} The proof of (i) follows immediately from the antisymmetry of the flux $\Psi_{jk}^n$. To prove (ii), we notice that when $\Psi_{jk}^n>0$, it takes the value $\rho_j^n$ and when $\Psi_{jk}^n<0$, it takes the value $- \rho_k^n$. Then, we have
\begin{eqnarray*}
&&\hspace{-1cm}
\rho_j^{n+1} = \rho_j^n - J \Delta t \Big( \sum_{k \in {\mathcal N}_j, \,  \Psi_{jk}^n >0} \rho_j^n - \sum_{k \in {\mathcal N}_j, \, \Psi_{jk}^n <0} \rho_k^n \Big) \\
&&\hspace{1cm}
= \rho_j^n \Big( 1 - J \Delta t  \sum_{k \in {\mathcal N}_j, \,  \Psi_{jk}^n >0} 1 \Big) + J \Delta t \sum_{k \in {\mathcal N}_j, \,  \Psi_{jk}^n <0} \rho_k^n \\
\end{eqnarray*}
Now, since $\sum_{k \in {\mathcal N}_j, \Psi_{jk}^n >0} 1 \leq d_j$, the first term is nonnegative under the CFL condition. The second term is nonnegative by assumption. This ends the proof. \endproof

\subsection{A simple cellular automaton on networks}
\label{sub:master_cell_master}

Again, like in section \ref{sub:master_cell}, we first consider the case where the rates $\gamma_j^n$ are independent of the node-densities $(\rho_j^n)_{j \in {\mathcal J}}$. Then, the node-densities can be ignored. The random variables consist of the node states $z_j$ for $j \in {\mathcal J}$ and the discrete state-space is given by $\Sigma^J = \prod_{j \in {\mathcal J}} {\mathcal N}_j$. We denote by $\vec{z}=(z_j)_{j \in {\mathcal J}}$ an element of $\Sigma^J$. A measure $\phi$ on $\Sigma^J$ is defined like in section \ref{sub:master_cell} by 
\begin{eqnarray}
&&\hspace{-1cm}
\langle \phi, \varphi \rangle_{\Sigma^J} :=  \sum_{\vec{z} \in \Sigma^J}  \phi(\vec{z}) \, \varphi(\vec{z}).
\label{eq:disc_dual_network}
\end{eqnarray}
The probability distribution function (pdf) of $\vec{z}$ at time $t_n$ is still denoted by  ${\mathcal{F}}^n(\vec{z})$. Let $\varphi$ be a smooth test function on $\Sigma^J$ with values in ${\mathbb R}$. As before, the expectation of the random variable $\varphi (\vec{z}^n)$ for all realizations of $\vec{z}^n$ with distribution ${\mathcal{F}}^n$ is given by (\ref{eq:expect}) (with $N$ replaced by $J$). We define the Markov transition operator $Q$ from state $\vec{z}^n$ to $\vec{z}^{n+1}$ by (\ref{eq:Qphi_def}) and we get (\ref{eq:meanfieldderv_1}) (again with $N$ replaced by $J$). In the limit $J \Delta t \to 0$, with $n J \Delta t \to t$, we have ${\mathcal F}^{n} (\vec z) \to {\mathcal F} (\vec z,t)$ where ${\mathcal F} (\vec z,t)$ satisfies the time-continuous master equation (\ref{eq:meanfieldderv_2}) associated to the adjoint operator $Q^*$ to $Q$. It is written 
\begin{eqnarray}
\label{eq:meanfield_network}
\frac{\partial {\mathcal F}}{\partial t} ( \vec{z},t ) = L^* {\mathcal F}( \vec{z},t ), 
\end{eqnarray}
with 
\begin{eqnarray}
\label{eq:L*_network}
L^* =  \lim_{J \Delta t \to 0} \frac{1}{J \Delta t}  (Q^* - \mbox{Id}) \quad \quad \mbox{and} \quad \quad Q \varphi (\vec{z}) = {\mathbb E} \left\{ \varphi(\vec{z}^{n+1})  \, \big| \,  \vec{z}^n = \vec{z} \right\}. 
\end{eqnarray}
We write the master equation explictly in the next proposition:

\begin{proposition}
The master equation for the time-continuous version of the CA described above when the rates $\gamma_j$ are independent of the node-densities $(\rho_j)_{j \in {\mathcal J}}$ is given by 
\begin{equation}
\frac{\partial}{\partial t}  {\mathcal{F}} (\vec{z}, t)  =  \frac{1}{J} \sum_{j \in {\mathcal J}} \frac{1}{d_j-1}  \sum_{z_j' \in {\mathcal N}_j\setminus \{z_j\}}  \big( \, \gamma_j(z_j', \hat z_j) \mathcal{F}(z_j', \hat z_j, t)  - \gamma_j(z_j, \hat z_j) \mathcal{F}(z_j, \hat z_j, t) \,  \big),    
\label{eq:master_CA_net}
\end{equation}
where we denote by $\hat z_j$ the vector of length $J-1$ collecting all node states but that corresponding to node $j$ and by $(z_j', \hat z_j)$ a state vector where the state of the $j$-th node is $z_j' \in {\mathcal N}_j\setminus \{z_j\}$ and the states of the other nodes are given by $\hat z_j$. 
\label{prop:master_CA_net}
\end{proposition}

This equation has a similar form and meaning as (\ref{eq:master_CA}) (except that now more than 2 nodes may be connected to a given node) and we refer to the paragraph following Prop. \ref{prop:master_CA} for its interpretation. 

\medskip
{\bf Proof of Prop. \ref{prop:master_CA_net}.} The proof follows the same strategy as that of Prop. \ref{prop:master_CA}. Let $\varphi$ be a smooth test function. Again the probability that a given $k$-tuple of cells switch states is ${\mathcal O}\big( (J \Delta t)^k\big)$. Therefore, the probability that there are strictly more than one change is ${\mathcal O} \big( (J\Delta t)^2 \big)$ while that of only one change is ${\mathcal O}(J \Delta t)$. This leads to:  
\begin{eqnarray*}
&&\hspace{-0.5cm} 
 \langle (Q^* - \mbox{Id}) {\mathcal F}^{n}, \varphi \rangle_{\Sigma^J}
= {\mathbb E} \left\{ {\mathbb E} \left\{ \varphi(\vec{z}^{n+1})- \varphi(\vec{z}^{n}) \, \big| \,  \vec{z}^n \right\}  \vec{z}^n \right\} \nonumber\\
&&\hspace{-0.5cm} 
=   \Big\langle  \,  \mathcal{F}^{n}(\vec{z}) \, , \,  \sum_{j \in {\mathcal J}} \,  \frac{1}{d_j-1} \sum_{z_j' \in {\mathcal N}_j \setminus \{ z_j\}} \, \big( \, \varphi(z_j', \hat z_j)  \, - \, \varphi(z_j, \hat z_j) \, \big) \,   \nonumber \\
&&\hspace{5cm} 
(1 - e^{- \gamma_j(z_j, \hat z_j)\Delta t}) \,  \prod_{i \not = j} e^{- \gamma_i(z_j, \hat z_j) \Delta t} \, \Big\rangle_{\Sigma^J} \, + \, {\mathcal O} \big( (J \Delta t)^{2} \big)
\end{eqnarray*}
Using (\ref{eq:disc_dual_network}), Taylor expansion when $ J \Delta t \ll 1$ and the fact that the restriction to ${\mathcal N}_j \setminus \{ z_j\}$ in the second sum can be removed since the added term is simply zero, we get: 
\begin{eqnarray*}
&&\hspace{-0.5cm} 
 \langle (Q^* - \mbox{Id}) {\mathcal F}^{n}, \varphi \rangle_{\Sigma^J}
=   \Delta t \, \sum_{\vec{z} \in \Sigma^J}  \, \sum_{j \in {\mathcal J}} \,  \frac{1}{d_j-1} \sum_{z_j' \in {\mathcal N}_j} \, \big( \, \varphi(z_j', \hat z_j)  \, - \, \varphi(z_j, \hat z_j) \, \big) \,  \nonumber \\
&&\hspace{9cm} 
 \, \gamma_j(z_j, \hat z_j) \, \,  \mathcal{F}^{n}(z_j, \hat z_j) \, + \,  {\mathcal O} \big( (J \Delta t^{2}) \big) .
\end{eqnarray*}
Now, pulling the summation over $j$ out and decomposing the summation over $\vec{z}$ in a summation over $\hat z_j$ and a summation over $z_j$, we get: 
\begin{eqnarray*}
&&\hspace{-0.5cm} 
 \langle (Q^* - \mbox{Id}) {\mathcal F}^{n}, \varphi \rangle_{\Sigma^J}
=  \Delta t \,  \sum_{j \in {\mathcal J}} \,  \frac{1}{d_j-1} \,  \sum_{\hat z_j \in \Sigma^{J\setminus\{j\}}} \sum_{z_j \in {\mathcal N}_j}  \,   \sum_{z_j' \in {\mathcal N}_j} \, \big( \, \varphi(z_j', \hat z_j)  \, - \, \varphi(z_j, \hat z_j) \, \big)   \nonumber \\
&&\hspace{9cm} 
\gamma_j(z_j, \hat z_j) \,   \,  \mathcal{F}^{n}(z_j, \hat z_j) \, + \, {\mathcal O} \big( (J \Delta t^{2}) \big) . 
\end{eqnarray*}
We can now exchange $z_j$ and $z_j'$ in the first term and obtain: 
\begin{eqnarray*}
&&\hspace{-0.5cm} 
 \langle (Q^* - \mbox{Id}) {\mathcal F}^{n}, \varphi \rangle_{\Sigma^J}
=  \Delta t \,  \sum_{j \in {\mathcal J}} \,  \frac{1}{d_j-1} \,  \sum_{\hat z_j \in \Sigma^{J\setminus\{j\}}} \sum_{z_j \in {\mathcal N}_j}  \,   \sum_{z_j' \in {\mathcal N}_j} \, \varphi(z_j, \hat z_j)  \nonumber \\
&&\hspace{4cm} 
\big( \, \gamma_j(z_j', \hat z_j) \,   \,  \mathcal{F}^{n}(z_j', \hat z_j)  \, - \, \,  \gamma_j(z_j, \hat z_j) \,   \,  \mathcal{F}^{n}(z_j, \hat z_j) \big) \, + \, {\mathcal O} \big( (J \Delta t^{2}) \big) . 
\end{eqnarray*}
Collecting the summation over $\hat z_j$ and over $z_j$ into a summation over $\vec{z}$, pulling this summation out and using again 
(\ref{eq:disc_dual_network}), we finally find:
\begin{eqnarray*}
&&\hspace{-1cm} 
 \langle (Q^* - \mbox{Id}) {\mathcal F}^{n}, \varphi \rangle_{\Sigma^J}
\\
&&\hspace{-1cm} 
= \Delta t \, \Big\langle \sum_{j \in {\mathcal J}} \,  \frac{1}{d_j-1} \,   \sum_{z_j' \in {\mathcal N}_j} \, \big( \, \gamma_j(z_j', \hat z_j) \,   \,  \mathcal{F}^{n}(z_j', \hat z_j)  \, - \, \,  \gamma_j(z_j, \hat z_j) \,   \,  \mathcal{F}^{n}(z_j, \hat z_j) \big) \, ,  \, \varphi(\vec{z}) \Big\rangle_{\Sigma^J} \nonumber \\
&&\hspace{12cm} 
 + \, {\mathcal O} \big( (J \Delta t^{2}) \big) . 
\end{eqnarray*}
which ends the proof. \endproof

In the next section, we consider the case where the rates $\gamma_j^n$ depend on the node-densities $(\rho_j^n)_{j \in {\mathcal J}}$.

\subsection{The master equation for the sweeping process on networks}
\label{sub:master_dens_net}

We now consider the full sweeping process on the network as described above. 
The random variables are now the node states $\vec{z} \in \Sigma^J$ and the node densities $\vec{\rho}=(\rho_j)_{j \in {\mathcal J}}$ with  $\rho_j \in {\mathbb R}_+$. The discrete state space for  $J$ nodes is therefore $\mathbb A^J = \Sigma^J  \times {\mathbb R}_+^J$. A measure $\phi$ on $\mathbb A^J$ is defined by its action on a continuous function $\varphi$ on $\mathbb A^J$ by: 
\begin{eqnarray*}
\langle \phi , \varphi \rangle_{{\mathbb A}^J} = \sum_{\vec{z} \in \Sigma^J}  \int_{{\mathbb R}_+^J}  \phi(\vec{z},\vec{\rho}) \, \varphi(\vec{z},\vec{\rho}) \, d \vec{\rho}.
\end{eqnarray*}
We also denote $\gamma_j=\gamma_j(\vec{z},\vec{\rho})$ and $\vec{\Xi}^n = (\xi_j^n)_{j \in {\mathcal J}}$, with 
$$ \xi_j^n = \sum_{k \in {\mathcal N}_j} \Psi_{jk}^n. $$
Then, the vector version of the density update is 
\begin{equation} 
\vec{\rho}^{n+1} - \vec{\rho}^n + J \Delta t \, \vec{\Xi}^n =0. 
\label{eq:rho_update_net}
\end{equation}
In the limit $J \Delta t \to 0$, we have the following proposition, whose proof is identical to that of Proposition \ref{prop:master_full} and is left to the reader.

\begin{proposition}
The master equation for the time-continuous version of the sweeping process described above when the rates $\gamma_j$ depend on both the node-states $\vec{z}$ and densities $\vec{\rho}$ is given by 
\begin{eqnarray}
&&\hspace{-1.5cm}
\Big( \frac{\partial}{\partial t}  {\mathcal{F}}  
-  \nabla_{\vec \rho} \cdot \big( \vec{\Xi} \, {\mathcal F} \big) \Big) (\vec{z}, {\vec \rho}, t)   \nonumber \\
&&\hspace{-1cm}
=   \frac{1}{J} \sum_{j \in {\mathcal J}} \frac{1}{d_j-1}  \sum_{z_j' \in {\mathcal N}_j\setminus \{z_j\}}  \big( \, \gamma_j(z_j', \hat z_j, \vec{\rho}) \mathcal{F}(z_j', \hat z_j, \vec{\rho}, t)  - \gamma_j(z_j, \hat z_j, \vec{\rho}) \mathcal{F}(z_j, \hat z_j, \vec{\rho},  t) \,  \big),    
\label{eq:master_net}
\end{eqnarray}
in strong form or 
\begin{eqnarray}
&&\hspace{-1cm}
\langle \frac{\partial {\mathcal F}}{\partial t}, \varphi \rangle_{{\mathbb A}^J}
=  - \langle {\mathcal F} , \nabla_{\vec \rho} \varphi \cdot \vec{\Xi} \, \rangle_{{\mathbb A}^J} \nonumber \\
&&\hspace{0cm}
+  \big\langle   \mathcal{F}(\vec{z})\, , \, \frac{1}{J} \sum_{j \in {\mathcal J}} \,  \frac{1}{d_j-1} \gamma_j(z_j, \hat z_j, \vec{\rho}) \,\sum_{z_j' \in {\mathcal N}_j} \, \big( \, \varphi(z_j', \hat z_j, \vec{\rho})  \, - \, \varphi(z_j, \hat z_j, \vec{\rho}) \, \big)
 \,  \,   \big\rangle_{{\mathbb A}^J} , 
\label{eq:master_net_2}
\end{eqnarray}
for any smooth test function $\varphi$ on ${\mathbb A}^J$ with values in ${\mathbb R}$, in weak form, where we recall that 
\begin{eqnarray}
&&\hspace{-1cm}
\vec{\Xi} = (\xi_j)_{j \in {\mathcal J}}, \quad \mbox{ with }  \quad  
\xi_j = \sum_{k \in {\mathcal N}_j} \Psi_{jk},  
\label{eq:Xi} \\
&&\hspace{-1cm}
\Psi_{jk} =  \rho_j \delta_{z_j \, k} - \rho_k \delta_{z_k \, j}.  
\label{eq:flux_network_2}
\end{eqnarray}
We have noted $ \nabla_{\vec \rho} \varphi \cdot \vec g = \sum_{j \in {\mathcal J}} g_j \partial_{\rho_j} \varphi $ and $\nabla_{\vec \rho} \cdot \vec g \, \varphi =  \varphi \sum_{j \in {\mathcal J}} \partial_{\rho_j} g_j$ for any functions $\varphi(\vec\rho)$ and $\vec g (\vec \rho) = (g_j(\vec \rho))_{j \in {\mathcal J}}$. 
\label{prop:master_full_net}
\end{proposition}

Again, the form and meaning of (\ref{eq:master_net}) is the same as that of (\ref{eq:master8}) and we refer to the paragraph following Prop. \ref{prop:master_full} for its interpretation. The only remark worth being made is that now, the total flux $\xi_j$ at node $j$ does not take the form of a simple difference of neighboring fluxes, like in (\ref{eq:master8}), but has the more complex expression (\ref{eq:Xi}). However, it is readily seen that this expression reduces to $\vec{\Psi}_+ - \vec{\Psi}_-$ in the one-dimensional case.

\subsection{Single particle closure for networks}
\label{sub:single_net}

The goal of this section is again to compute a closed system of equations for the one-particle marginals of ${\mathcal{F}}(\vec{z},\vec{\rho},t)$. We define the marginals according to: 

\begin{definition}
For any $j \in {\mathcal J}$, we define the marginal density ${\bf f}_j$ on ${\mathbb A}$ by duality by 
\begin{equation}
\langle {\bf f}_j(z_j, \rho_j, t) , \varphi_j(z_j, \rho_j) \rangle_{{\mathbb A}^J} = \langle {\mathcal F}(z_j, \hat z_j, \rho_j, \hat \rho_j ,t) , \varphi_j (z_j, \rho_j) \rangle_{{\mathbb A}^J} , 
\label{eq:marginal_weak_net}
\end{equation}
where $\hat \rho_j$ is a $J-1$-dimensional vector collecting all $\rho_m$ for $m \in {\mathcal J}$, with $m \not = j$, and with any test function $\varphi_j(z_j,\rho_j)$ of the single variables $(z_j,\rho_j)$. Equivalently, we have: 
\begin{equation}
{\bf f}_j(z_j, \rho_j, t) = \langle {\mathcal F}(z_j, \hat z_j, \rho_j, \hat \rho_j ,t) , 1 \rangle_{\hat {\mathbb A}_j} , 
\label{eq:marginal_strong_net}
\end{equation}
where $\langle \cdot , \cdot \rangle_{\hat {\mathbb A}_j}$ denotes the duality between measures and functions of the variables $(\hat z_j, \hat \rho_j)$ in ${\mathbb A}^{J-1}$ (and ${\mathbb A}^{J-1}$ is denoted by $\hat {\mathbb A}_j$ when such a duality is considered). 
\label{def:marginal_net}
\end{definition}

To get an equation for ${\bf f}_j$, we use the master equation in weak form (\ref{eq:master_net_2}) with a test function $\varphi_j(z_j,\rho_j)$ of the single variables $(z_j,\rho_j)$. The resulting equation is given by the following proposition:

\begin{proposition}
Define:
\begin{eqnarray}
&&\hspace{-1cm}
\overline{\xi_j}(t) \, {\bf f}_j(t) := \langle {\mathcal F}(t) , \xi_j \rangle_{\hat {\mathbb A}_j} , \label{eq:effective_flux_net}\\
&&\hspace{-1cm}
\bar \gamma_j(t) \,  {\bf f}_j (t):= \langle {\mathcal F}(t) , \gamma_j \rangle_{\hat {\mathbb A}_j} . \label{eq:effective_rate_net}
\end{eqnarray}
The functions $\overline{\xi_j}(t)$ and $\bar \gamma_j(t)$ are functions of $(z_j, \rho_j)$ only. Then, the equation for the marginal ${\bf f}_j$ is written in weak form: 
\begin{eqnarray}
&&\hspace{-1cm}
\langle \frac{\partial {\bf f}_j}{\partial t}, \varphi_j \rangle_{{\mathbb A}_j} 
=  - \langle {\bf f}_j , \overline{\xi_j}(t) \,\partial_{\rho_j} \varphi_j  \, \rangle_{{\mathbb A}_j} \nonumber \\
&&\hspace{1cm}
+  \frac{1}{J} \,  \frac{1}{d_j-1} \langle \,  {\bf f}_j ,  \bar \gamma_j(z_j, \rho_j,t) \,\sum_{z_j' \in {\mathcal N}_j} \,   \big( \varphi_j(z_j', \rho_j) - \varphi_j(z_j, \rho_j) \big) \, \big\rangle_{{\mathbb A}_j} , 
\label{eq:marginal1_net}
\end{eqnarray}
where ${\mathbb A}_j = {\mathcal N}_j \times {\mathbb R}_+$, and in strong form
\begin{eqnarray}
&&\hspace{-1cm}
\Big( \frac{\partial}{\partial t}  {\bf f}_j 
-  \partial_{\rho_j} \big( \overline{\xi_j}(t) \, {\bf f}_j  \big) \Big) (z_j, \rho_j, t)
  \nonumber \\
&&\hspace{1cm}
= \frac{1}{J}  \,  \frac{1}{d_j-1} \sum_{z_j' \in {\mathcal N}_j} \, \big( \, \bar \gamma_j (z_j', \rho_j,t) {\bf f}_j (z_j', \rho_j, t) 
- \bar \gamma_j(z_j, \rho_j, t) {\bf f}_j (z_j, \rho_j, t) \, \big) .
\label{eq:marginal2_net}
\end{eqnarray}
\label{prop:marginal_eq_net}
\end{proposition}

We now make the propagation of chaos assumption, which in the network framework reads as follows: 

\begin{assumption}
We assume that the joint pdf ${\mathcal F}(\vec z, \vec \rho, t)$ is written as:
\begin{eqnarray}
&&\hspace{-1.3cm}
{\mathcal F}(\vec z, \vec \rho, t) = \prod_{j \in {\mathcal J}} {\bf f}_j(z_j, \rho_j, t).
\label{eq:chaos_net}
\end{eqnarray}
\label{ass:chaos_net}
\end{assumption}

With this assumption, we can simplify the expressions of the flux (\ref{eq:effective_flux_net}). We have the following: 

\begin{lemma}
Under the chaos assumption (Assumption \ref{ass:chaos_net}), the flux (\ref{eq:effective_flux_net}) is given by:
\begin{eqnarray}
&&\hspace{-1cm}
\overline{\xi_j}(z_j,\rho_j,t) = \rho_j  - \sum_{k \in {\mathcal N}_j} \int_0^\infty \rho_k \, {\bf f}_k (j, \rho_k, t) \,  d \rho_k
\label{eq:eff_flux_chaos_net} 
\end{eqnarray}
\label{lem:flux_rates_net}
\end{lemma}

\medskip
\noindent
{\bf Proof.} From equation (\ref{eq:effective_flux_net}), we have 
\begin{eqnarray*}
&&\hspace{-1cm}
\overline{\xi_j}(z_j, \rho_j, t) \, {\bf f}_j(z_j, \rho_j, t) = \langle \prod_{\ell \in {\mathcal J}} {\bf f}_\ell(z_\ell, \rho_\ell, t) , \xi_j \rangle_{\hat {\mathbb A}_j} , 
\end{eqnarray*}
Inserting (\ref{eq:Xi}), (\ref{eq:flux_network_2}) into this equation leads to 
\begin{eqnarray*}
&&\hspace{-1cm}
\overline{\xi_j}(z_j, \rho_j, t) =  \langle \prod_{\ell  \in {\mathcal J}, \, \ell \not = j} {\bf f}_\ell(z_\ell, \rho_\ell, t) , \sum_{k \in {\mathcal N}_j} (\rho_j \delta_{z_j \, k} - \rho_k \delta_{z_k \, j}) \rangle_{\hat {\mathbb A}_j} . 
\end{eqnarray*}
Interchanging the summation over $k$ and over $\hat {\mathbb A}_j$, we get:
\begin{eqnarray}
&&\hspace{1cm}
\overline{\xi_j}(z_j, \rho_j, t) =\sum_{k \in {\mathcal N}_j} \langle \prod_{\ell  \in {\mathcal J}, \, \ell \not = j} {\bf f}_\ell(z_\ell, \rho_\ell, t) ,  (\rho_j \delta_{z_j \, k} - \rho_k \delta_{z_k \, j}) \rangle_{\hat {\mathbb A}_j} . 
\label{eq:chaos_calc}
\end{eqnarray}

The term $\rho_j \delta_{z_j \, k}$ only depends on the state of the $j$-th node. Therefore, it can be taken out of the bracket over $\hat {\mathbb A}_j$. There remains $\langle \prod_{\ell  \in {\mathcal J}}, {\bf f}_\ell(z_\ell, \rho_\ell, t) , \, 1 \rangle_{\hat {\mathbb A}_j}$ which is equal to $1$ because each ${\bf f}_\ell$ is a probability density. Therefore, the positive term at the right-hand side of (\ref{eq:chaos_calc}) reduces to $\rho_j \, \sum_{k \in {\mathcal N}_j} \delta_{z_j \, k}$. Since, there is only one node $k \in {\mathcal N}_j$ such that the state $z_j$ of node $j$ is equal to $k$, we have $\sum_{k \in {\mathcal N}_j} \delta_{z_j \, k} = 1$. Finally, the production term reduces to $\rho_j$. 
 
The expression of the negative term at the right-hand side of (\ref{eq:chaos_calc}), follows from the fact that   
\begin{eqnarray*}
&&\hspace{-1cm}
\sum_{k \in {\mathcal N}_j} \langle \prod_{\ell  \in {\mathcal J}, \, \ell \not = j} {\bf f}_\ell(z_\ell, \rho_\ell, t) , \rho_k \delta_{z_k \, j} \rangle_{\hat {\mathbb A}_j} =  \\
&&\hspace{1cm}
= \sum_{k \in {\mathcal N}_j} \sum_{\hat z_j \in \Sigma^{J-1}} \int_{\hat \rho_j \in {\mathbb R}_+^{J-1}} \rho_k \delta_{z_k j} \prod_{\ell \not = j, \ell \in {\mathcal{J} } } {\bf f}_\ell(z_\ell, \rho_\ell, t) \, d \hat \rho_j 
\end{eqnarray*}
In the previous formula, only the sum over $z_k$ and integral over $\rho_k$ is different from $1$ because again, each ${\bf f}_\ell$ is a probability on the state space $(z_\ell, \rho_\ell)$. Now, because of the muliplication by $\delta_{z_k j}$, the sum over $z_k$ has only one non-zero contribution, that corresponding to $z_k = j$. The resulting value of the sink term is therefore equal to 
\begin{eqnarray*}
&&\hspace{-1cm}
\sum_{k \in {\mathcal N}_j} \langle \prod_{\ell  \in {\mathcal J}, \, \ell \not = j} {\bf f}_\ell(z_\ell, \rho_\ell, t) , \rho_k \delta_{z_k \, j} \rangle_{\hat {\mathbb A}_j} =
\sum_{k \in {\mathcal N}_j} \int_{\rho_k \in {\mathbb R}_+} \rho_k \, {\bf f}_k(j, \rho_k, t) \, d\rho_k
\end{eqnarray*}
Collecting the calculations of the production and sink terms, we are led to (\ref{eq:eff_flux_chaos_net}), which ends the proof. \endproof

Now collecting the results of Proposition \ref{prop:marginal_eq_net} and Lemma \ref{lem:flux_rates_net}, we can state the following theorem:

\begin{theorem}
Under the closure assumption (\ref{eq:chaos_net}), the equation for the marginal ${\bf f}_j$ is written in weak form: 
\begin{eqnarray}
&&\hspace{-1cm}
\langle \frac{\partial {\bf f}_j}{\partial t}, \varphi_j \rangle_{{\mathbb A}_j} 
=  - \langle {\bf f}_j , \big( \rho_j  - \sum_{k \in {\mathcal N}_j} \int_0^\infty \rho_k \, {\bf f}_k (j, \rho_k, t) \,  d \rho_k \big)  \,\partial_{\rho_j} \varphi_j  \, \rangle_{{\mathbb A}_j} \nonumber \\
&&\hspace{1cm}
+  \frac{1}{J} \,  \frac{1}{d_j-1} \langle \,  {\bf f}_j ,  \bar \gamma_j(z_j, \rho_j,t) \,\sum_{z_j' \in {\mathcal N}_j} \,   \big( \varphi_j(z_j', \rho_j) - \varphi_j(z_j, \rho_j) \big) \, \big\rangle_{{\mathbb A}_j} , 
\label{eq:marginal3_net}
\end{eqnarray}
where ${\mathbb A}_j = {\mathcal N}_j \times {\mathbb R}_+$, and in strong form
\begin{eqnarray}
&&\hspace{-1cm}
\Big( \frac{\partial}{\partial t}  {\bf f}_j 
-  \partial_{\rho_j} \big( \big( \rho_j  - \sum_{k \in {\mathcal N}_j} \int_0^\infty \rho_k \, {\bf f}_k (j, \rho_k, t) \,  d \rho_k \big) \, {\bf f}_j  \big) \Big) (z_j, \rho_j, t)
  \nonumber \\
&&\hspace{1cm}
= \frac{1}{J}  \,  \frac{1}{d_j-1} \sum_{z_j' \in {\mathcal N}_j} \, \big( \, \bar \gamma_j (z_j', \rho_j,t) {\bf f}_j (z_j', \rho_j, t) 
- \bar \gamma_j(z_j, \rho_j, t) {\bf f}_j (z_j, \rho_j, t) \, \big) .
\label{eq:marginal4_net}
\end{eqnarray}
Here, $\bar \gamma_j(z_j, \rho_j, t)$ is given by 
\begin{eqnarray}
&&\hspace{-1cm}
\bar \gamma_j(t) \,  = \langle \prod_{\ell \in {\mathcal J}, \, \ell \not = j} {\bf f}_\ell (z_\ell, \rho_\ell, t) , \gamma_j \rangle_{\hat {\mathbb A}_j} . \label{eq:effective_rate_2_net}
\end{eqnarray}
\label{thm:marginal_eq_net}
\end{theorem}

Eq. (\ref{eq:marginal4_net}) provides the evolution of the $1$-node pdf in the phase space consisting of the $j$-th cell state space ${\mathcal N}_j$ for $z_j$ and the density space ${\mathbb R}_+$ for $\rho_j$. It takes the form of a transport equation in the $\rho_j$ variable (the left-hand side), with a collision term describing the rate of change of the $j$-th cell states $z_j$ (the right-hand side). The collision operator has a similar form and meaning as the right-hand side of Eq. (\ref{eq:master_CA}) or (\ref{eq:master_CA_net}) (but for the passage from the $J$-node pdf to the $1$-node pdf) and we refer to the paragraph following Prop. \ref{prop:master_CA} for its interpretation. The interesting feature in (\ref{eq:marginal4_net}) is the transport operator. Indeed, the flux term (inside the $\partial_{\rho_j}$ derivative) in the $j$-th cell pdf is given in terms of the average density in neighboring cells. This average density is obtained through integrating the neighboring cell pdf $f_k$ over the density variable $\rho_k$. Therefore, the various pdf are coupled altogether by this flux term in an integral fashion. To some extent, this coupling resembles a mean-field coupling like in Vlasov-type models. Another source of coupling of the various $1$-cell pdf is through the evaluation of the switching rates $\bar \gamma_j$, which depend on the pdf of some of the neighboring cells through the mean-field evaluation (\ref{eq:effective_rate_2_net}).

\setcounter{equation}{0}
\section{Summary and perspectives}
\label{sec:conclu}

We present a derivation of macroscopic equations for the large--time behavior of microscopic sweeping  processes coupled to density evolutions. Within the derivation a general master equation is considered and under a meanfield assumption kinetic equations are derived. We applied the general calculus to an example of pedestrian flow in small  corridors. An extension of the ideas towards flows on networks has also been presented. 

In future work we discuss equations arising from  a Chapman--Enskog like expansion for the cell-width going to zero.  Further, it would be interesting to analyze a Taylor expansion of the rate equation  (\ref{eq:eff_rate_largeN}) for strongly confined kernels $w$. Another open problem is the combination of the meanfield assumption and kernel localized within a finite number of cells (such as e.g. a nearest neighbor interaction) leading to possibly correlated particle distributions. Physically more sophisticated CA may be envisionned. For instance, we could introduce different particle densities for left and right going particles, and according to the state of the cell, move one of the population while the other population stays immobile. Other improvements would consist of taking into account finite network capacity or more generally, more complex rules for the computation of the switching rates. For instance, time delays could be introduced to model the finiteness of the information propagation speed. Finally, the hypotheses made here, i.e. propagation of chaos and mean-field limit need to be validated by intensive numerical simulations.


\end{document}